\documentclass[12pt]{article}
\pdfoutput=1
\usepackage{graphicx}
\usepackage{color,amsmath,amssymb,exscale,psfrag,epsfig}
\usepackage{cite,color,url}
\usepackage{pdfpages}
\usepackage{multirow}
\usepackage[colorlinks=true
,urlcolor=blue
,anchorcolor=blue
,citecolor=blue
,filecolor=blue
,linkcolor=blue
,menucolor=blue
,linktocpage=true
,pdfproducer=medialab
]{hyperref}
\input epsf
\usepackage{lineno}

\newcommand{\m}[1]{\marginpar{{\tiny *}} }

\newcommand{\Dslash}{{\not \!\!D}}

\def\bea{\begin{eqnarray}}
\def\eea{\end{eqnarray}}

\newcommand\f[2]{\frac{#1}{#2}}
\newcommand{\itsz}{\it}
\catcode`\@=11
\def\lsim{\mathrel{\mathpalette\@versim<}}
\def\gsim{\mathrel{\mathpalette\@versim>}}
\def\@versim#1#2{\vcenter{\offinterlineskip
\ialign{$\m@th#1\hfil##\hfil$\crcr#2\crcr\sim\crcr } }}
\catcode`\@=12
\catcode`@=12
\begin{document}
\thispagestyle{empty}
\begin{flushright}
ICAS 23/16 \\
ZU-TH 38/16
\end{flushright}
\begin{center}
{\Large \bf Graviton resonance phenomenology \\ and a pNGB Higgs at the LHC} \\
\vspace{0.2in}
{\bf Ezequiel Alvarez$^{(a,b)}$,
Leandro Da Rold$^{(c)}$,\\[1ex]
Javier Mazzitelli$^{(a,d)}$
and
Alejandro Szynkman$^{(e)}$}
\vspace{0.1in} \\
{\sl $^{(a)}$ International Center for Advanced Studies (ICAS), UNSAM, Campus Miguelete \\
25 de Mayo y Francia, (1650) Buenos Aires, Argentina
}
\\[1ex]
{\sl $^{(b)}$ International Center for Theoretical Physics (ICTP), Strada Costiera, 11, Trieste, Italy 
}
\\[1ex]
{\sl $^{(c)}$ Centro At\'omico Bariloche, Instituto Balseiro and CONICET \\
Av. Bustillo 9500, 8400, S. C. de Bariloche, Argentina
}
\\[1ex]
{\sl $^{(d)}$ Physik-Institut, Universit\"at Z\"urich, \\ 
Winterthurerstrasse 190, CH-8057 Z\"urich, Switzerland
}
\\[1ex]
{\sl $^{(e)}$ IFLP, Dpto. de F\'{\i}sica, CONICET, UNLP \\ 
C.C. 67, 1900 La Plata, Argentina} \\
\end{center}

\begin{abstract}
We present an effective description of a spin two massive state and a pseudo Nambu-Goldstone boson Higgs in a two site model. Using this framework we model the spin two state as a massive graviton and we study its phenomenology at the LHC. We find that a reduced set of parameters can describe the most important features of this scenario. We address the question of which channel is the most sensitive to detect this graviton.  Instead of designing search strategies to estimate the significance in each channel, we compare the ratio of our theoretical predictions to the limits set by available experimental searches for all the decay channels and as a function of the free parameters in the model.  We discuss the phenomenological details contained in the outcome of this simple procedure.  The results indicate that, for the studied masses between 0.5 and 3 TeV,  the channels to look for such a graviton resonance are mainly $ZZ$, $WW$ and $\gamma\gamma$.  This is the case even though top and bottom quarks dominate the branching ratios, since their experimental sensitivity is not as good as the one of the electro-weak gauge bosons.  We find that as the graviton mass increases, the $ZZ$ and $WW$ channels become more important because of its relatively better enhancement over background, mainly due to fat jet techniques.  We determine the region of the parameter space that has already been excluded and the reach for the LHC next stages. We also estimate the size of the loop-induced contributions to the production and decay of the graviton, and show in which region of the parameter space their effects are relevant for our analysis.

\end{abstract}

\vspace*{1mm}
\noindent {\footnotesize E-mail:
{\tt \href{mailto:sequi@df.uba.ar}{sequi@df.uba.ar},
\href{mailto:jmazzi@physik.uzh.ch}{jmazzi@physik.uzh.ch},
\href{mailto:daroldl@cab.cnea.gov.ar}{daroldl@cab.cnea.gov.ar},\\
\href{mailto:szynkman@fisica.unlp.edu.ar}{szynkman@fisica.unlp.edu.ar}}}


\newpage
\section{Introduction}
After the discovery of a Higgs like state and the measurement of its properties, the main objective of the LHC is the search of new physics (NP). ATLAS and CMS have designed and conducted many different searches on new particles in the 8 and 13 TeV runs of the LHC. From the perspective of many theories beyond the Standard Model, neutral spin two massive states are one of the most attractive possibilities, due to its possible connection with gravity. ATLAS and CMS have searched for production of these states, designing search strategies for the different decay channels, as dibosons and pairs of fermions. The absence of positive signals has led to bounds on the cross sections of these channels in both runs. These bounds can be translated into limits in the masses and couplings of the massive spin two states. One of the goals of this paper is to analyze these bounds in a simple phenomenological model of a massive spin two state, as well as to determine the most sensitive channels where it could be detected. 

Massive spin two states appear naturally in theories where the Higgs is a composite state arising from a new strongly coupled sector. Usually, in composite Higgs models, besides the Higgs scalar, one can expect a whole bunch of new composite states. One of the most interesting examples are the colored composite partners of the top, that would cut off the large top contributions to the Higgs mass~\cite{Carena:2006bn,Contino:2006qr}. One can also expect partners of the others quarks and leptons, as well as spin one states associated to the gauge bosons and spin two states associated to the graviton. One realization of this scenario is achieved by considering the presence of a compact extra dimension~\cite{Randall:1999ee}. 
In this case the SM fields approximately correspond to the would-be zero-modes, and the composite resonances to the massive Kaluza-Klein (KK) states. The case of the Randall-Sundrum model with the SM fields in the bulk and the Higgs as the fifth component of a five dimensional gauge field is one of the most interesting possibilities~\cite{Agashe:2004rs}. Within this context, one of the most exciting phenomenological signals would be the single production at the LHC of the first resonance of the graviton, that would strongly support the solution of the hierarchy problem by new composite dynamics~\cite{Agashe:2007zd}.~\footnote{Another very interesting signal is the production of top partners, mainly by strong interactions~\cite{DeSimone:2012fs}.} 

In the present paper we will consider an effective description of the SM and the composite dynamics in the context of a two site model. This framework introduces the lowest layer of resonances and provides a very simple description, yet capturing the features important for the LHC phenomenology~\cite{Contino:2006nn}. We will describe the spin two state as a massive graviton of theory space~\cite{ArkaniHamed:2002sp}. We also consider the case where the Higgs is a pseudo Nambu-Goldstone boson (pNGB) arising from the strong dynamics~\cite{DeCurtis:2011yx,Panico:2011pw,Carena:2014ria,Panico:2015jxa}. Being the Higgs a composite state, we expect the massive graviton to couple strongly to the longitudinal components of the $W$ and $Z$. Since many searches have been optimized for diboson final states, for usual composite Higgs models the bounds are very stringent. We will show that if their couplings with the Higgs are suppressed, a huge volume of the parameter space with rather low graviton mass is still available. We will also show that in the case of a pNGB Higgs, these couplings can be parametrically suppressed. 

For the phenomenology of the massive graviton at the LHC we study the most sensitive channels and determine the allowed region of the parameter space. For this purpose we compare the production cross section of the massive graviton in the different decay channels with the present experimental sensitivity. In doing so one can incorporate many experimental aspects and have a quick understanding of the main phenomenological features of the graviton. It is straightforward to determine the dominant as well as the subdominant decay channels and to see how far from the present sensitivities these channels are. It is also very simple to determine the masses and couplings that are already excluded and to estimate the region of the parameter space that could be tested at the LHC with the present techniques and strategies. 

We also study a set of radiative corrections to the graviton couplings. We show that, even in the presence of many composite fermions arising form the extended symmetry of the new sector, the 1-loop corrections to the couplings with the gauge bosons are small and can be neglected for our analysis. An exception to this results is the case where the tree level couplings with the gauge boson are very small. In this case the 1-loop correction becomes important and can give interesting effects.

Our paper is organized as follows: in sec.~\ref{sec:model} we describe the models and compute the spectrum and couplings.  In sec.~\ref{sec:pheno} we perform the analysis of the graviton phenomenology at LHC. This section contains the most important results of the paper. In sec.~\ref{sec:loop} we estimate the 1-loop corrections and discuss their effects. We conclude in sec.~\ref{sec:conclusions}.

\section{Model}\label{sec:model}
We assume that, besides the SM, there is a new sector with a strong dynamics. The interactions of this sector generate bound states, including the Higgs boson and spin two states corresponding to massive gravitons. The masses of the composite states are of order TeV and the interactions between these states are assumed to be perturbative but still larger than one. The elementary gauge fields of the SM interact with the composite sector by weakly gauging some global symmetries. The elementary fermions of the SM have linear interactions with the composite sector, realizing partial compositeness of the SM fermions. We closely follow the description of the model given by us in Ref.~\cite{Alvarez:2016uzm}.

\begin{figure}[t] 
\begin{center}
\includegraphics[width=0.6\textwidth]{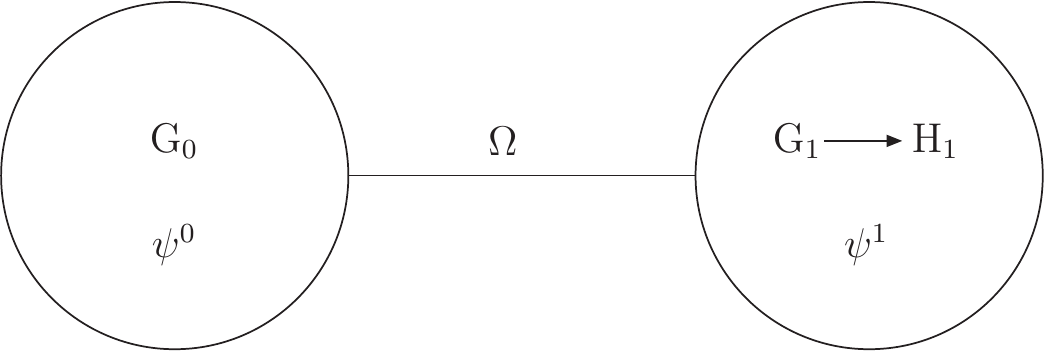}
\end{center}
\caption{\small Moose diagram describing the model in terms of a two site theory. Site-0 and site-1 have gauge symmetry groups G$_0$ and G$_1$. The Higgs arises from the spontaneous breaking of G$_1$ down to H$_1$ by the strong interactions on site-1. The link fields $\Omega$, transforming non-trivially under the symmetries on both sites, allow to connect the fields on site-0 and site-1.}
\label{fig-moose}
\end{figure}

A simplified description of the previous dynamics can be obtained by considering a two-site model, Fig.~\ref{fig-moose}. Site-0 is the {\itsz elementary} site, containing the SM gauge fields, the SM fermions and a massless graviton. Site-1 is the {\itsz composite} site, it contains only the first layer of resonances of the strongly interacting sector, one of these resonances corresponding to the Higgs doublet. An effective description of the resonances can be obtained by assuming that site-1 has a gauge symmetry G$_1$. We will take G$_1$ larger than the SM gauge symmetry, such that site-1 delivers a Higgs as a pNGB arising from the spontaneous breaking of G$_1$ to a smaller group $H_1$, with $H_1$ containing the SM gauge symmetry. G$_1$ can be taken to include also the custodial symmetry of the Higgs sector of the SM. We will give an explicit example in sec.~\ref{sec-higgs-sector}. On site-1 there is also a graviton associated to general coordinate transformations on that site. Besides there are vector-like fermions in representations of G$_1$, there is one multiplet of composite fermions for each multiplet of elementary fermions.\footnote{It is also possible to consider more than one composite fermion for each elementary one~\cite{Contino:2006qr,Csaki:2008zd,Andres:2015oqa}. For simplicity in this section we will consider that there is just one.} The gauge and Yukawa-like interactions between the fields on site-1 correspond to interactions between the resonances of the strongly coupled sector. Using $g_1$ to collectively denote the dimensionless couplings of the composite sector, we will assume $g_1$ weak but still larger than the elementary couplings: $1<g_1\ll 4\pi$. The mass scale on site-1 will be $f_1\sim$ TeV, and we will take the masses of the composite states $m_1\sim g_1 f_1$.

The elementary and composite sites are connected by link fields, that we will denote collectively as $\Omega$. There are link fields $U$ transforming under G$_0$ and G$_1$, that allow to write gauge invariant operators containing elementary and  composite fields. There is also a link field $Y_{10}$ that transforms under general coordinate transformations on both sites, and allows to compare fields located in the different sites. Denoting as $F_0$ and $F_1$ the general coordinate transformations on site-0 and site-1, then: $Y_{10}\to F^{-1}_1\circ Y_{10}\circ F_0$~\cite{ArkaniHamed:2002sp}. The main effect of these interactions, as can be shown in the unitary gauge where the maps correspond to the identity, is to generate a mass for some linear combinations of the fields on site-0 and site-1. We will show this in detail in~\ref{sec-mass-basis}.

The linear interactions between the elementary fermions and the strongly coupled sector leads to what is also known as partial compositeness of the fermions, since the mass eigenstates are linear superpositions of elementary and composite states. In the present model partial compositeness will be achieved by considering interactions linear in the  elementary and composite fermions, that will be connected by link fields to obtain gauge invariant operators. Since the composite fermions interact with the Higgs, in the unitary gauge and after electroweak symmetry breaking (EWSB), these mixing will generate masses for the SM fermions. We will assume that there are no bilinear interactions between the elementary fermions and the strongly coupled sector, thus masses of all the SM fermions are generated by partial compositeness.
 
As in Ref.~\cite{Contino:2006nn}, the Lagrangian of the model can be written as:
\begin{align}
&{\cal L}={\cal L}_0+{\cal L}_1+{\cal L}_{\rm mix} \ , \\
&{\cal L}_j={\cal L}_j^{\rm matter}+\sqrt{-G_j}2M_j^2R(G_j) + \dots\ , \qquad j=0,1 \ ,\label{eqLi}
\\
&{\cal L}_{\rm mix}={\cal L}_{\rm mix}^{\rm matter}+{\cal L}_{\rm mix}^{\rm grav} \ .
\end{align}
$G^j_{\mu\nu}$ and $M_j$ are respectively the metric and the scale of the gravitational interactions on site $j$, $M_0$ being of order $M_{\rm Pl}$ and $M_1$ of order TeV. In Eq.~(\ref{eqLi}) the dots allow for more terms, as a cosmological constant and a term for a dilaton. On each site there is a term with the Ricci scalar made from the corresponding metric, that contains a kinetic term for the gravitons. The Lagrangians ${\cal L}_j^{\rm matter}$ are:
\begin{equation}\label{eq-Lj}
{\cal L}_j^{\rm matter}=\sqrt{-G_j}\left[-\frac{1}{4g_j^2}F_j^{\mu\nu a}F_{j\mu\nu}^{a}+\bar\psi_j(i\Dslash-m_{\psi_j})\psi_j+\dots\right]\ .
\end{equation}
The fermion mass term and the dots are only present for site-1. ${\cal L}_0$ is similar to the SM Lagrangian, without the Higgs.
${\cal L}_1^{\rm matter}$ contains the kinetic terms of the gauge and the fermion fields on site-1, as well as the mass terms for the fermions of this site that are vector-like.
The dots stand for the terms of ${\cal L}_1^{\rm matter}$ describing a NGB Higgs on $G_1/H_1$, explicitly written in the second term of Eq.~(\ref{kinetic-ngb}), as well as Yukawa interactions whose explicit form depend on the symmetry groups and on the representations chosen for the fermions $\psi_1$. 

${\cal L}_{\rm mix}$ contains the terms mixing the fields of both sites. The mixing Lagrangian for the gravitational sector is shown in Eq.~(\ref{grav-mix}) in the unitary gauge~\cite{ArkaniHamed:2002sp}. ${\cal L}_{\rm mix}$ contains also terms mixing $\psi^0$ and $\psi^1$ that are shown in the first term of Eq.~(\ref{eq-ferm1}). Finally, the terms of ${\cal L}_{\rm mix}$ involving the gauge fields and the Higgs are shown in the first term of Eq.~(\ref{kinetic-ngb}), they will be discussed in detail in sec.~\ref{sec-higgs-sector}.

Of particular importance for the study of the phenomenology of the massive graviton are the interactions of the gravitons of both sectors. We will split the metrics in both sites into the Minkowski term and a fluctuation: $G^j_{\mu\nu}=\eta_{\mu\nu}+X^j_{\mu\nu}$. Expanding to linear order in the graviton fields we obtain:
\begin{equation}
{\cal L}_j\supset X^j_{\mu\nu}T_j^{\mu\nu} \ ,
\end{equation}
where the energy-momentum tensors on each site are defined as usual:
\begin{align}\label{eq-energymomentum-tensor}
T_j^{\mu\nu}=&-\frac{1}{g_j^2}F_j^{\mu\rho}F_{j\rho}^{\nu}+\eta^{\mu\nu}\frac{1}{4g_j^2}F_j^{\rho\sigma}F_{\rho\sigma}^j+\frac{i}{2}\bar\psi_j(\gamma^\mu D^\nu+\gamma^\nu D^\mu)\psi_j-\eta^{\mu\nu}\bar\psi_j(i\Dslash-m_j)\psi_j +\dots 
\end{align}
The dots stand for the contribution from the Higgs field, we will discuss its form and coupling in the following subsections.

\subsection{Gauge and Higgs sectors}\label{sec-higgs-sector}
Generically, since the coupling between the composite states are large: $g_1>1$, and the Higgs is the lightest composite state, we expect a composite graviton to decay copiously to Higgs pairs, and thus to longitudinal $W$ and $Z$ bosons. If the Higgs is completely localized on site-1, $X^1\to HH$ can easily overcome the bounds from direct searches. In the present model we will consider the Higgs arising as a pNGB, and we will show that in this case the decay to longitudinal EW gauge bosons can be naturally suppressed. In fact we will show that, in a simplified analysis of the graviton phenomenology at LHC, the factor suppressing this decay is one of the most important parameters for the description of the graviton phenomenology.

Let us start with the description of the Higgs sector by considering a specific example. Although the graviton phenomenology does not depend on the details of the pattern of symmetric breaking, we will show the well known example of SO(5)/SO(4). We choose G$_1$=SU(3)$_c\times$SO(5)$\times$U(1)$_X$ broken down to H$_1$=SU(3)$_c\times$SO(4)$\times$U(1)$_X$ by the strong dynamics. In this case the Higgs transforms as a ${\bf 4}_0$ of SO(4)$\times$U(1)$_X$, and it is color neutral. The extra U(1)$_X$ is required to obtain the proper hypercharge generator that is realized as $Y=T^{3R}+X$. From now on we will use $a$ and $\hat a$ to label the unbroken and broken generators of G$_1$, respectively.

It is convenient to extend spuriously the gauge symmetry on site-0 to G$_0$=SU(3)$_c\times$SO(5) $\times$ U(1)$_X$. This can be done by introducing non-dynamical fields that allow to furnish complete representations of the extended symmetry group.\footnote{We will choose the same representations of SO(5) for the fermions on site-0 and site-1.} Below we will use subindices 0 and 1 to specify the site to which the gauge symmetry belongs.

On site-1 there is a scalar field $U_1=e^{i\sqrt{2}\Pi_1/f_1}$ that transforms non-linearly under SO(5)$_1$: $U_1\to \hat g_1 U_1\hat h_1^\dagger$, with $\hat g_1\in$ SO(5)$_1$ and $\hat h_1\in$ SO(4)$_1$ depending on $\hat g_1$ and $\Pi$. This field parametrizes the spontaneous breaking SO(5)$_1$/SO(4)$_1$ at scale $f_1\sim$ TeV. As usual $\Pi_1$ can be written as: $\Pi_1=\Pi_1^{\hat a}T^{\hat a}$. There is another set of scalar fields $U_A=e^{i\sqrt{2}\Pi_A/f_A}$ that transform as: $U_A \to \hat g_0 U_A \hat g_1^\dagger$, with $\hat g_0$ and $\hat g_1$ elements of G$_0$ and G$_1$, respectively. $U_A$ parametrizes the breaking G$_0\times$G$_1$/G$_{0+1}$ at scale $f_A\sim$ TeV, with $\Pi_A=\Pi_A^rT^r$ and $T^r$ the broken generators. One can take $A$ to label the different groups in each site, such that there is one NGB field and one decay constant associated to each group: SU(3)$_c$, SO(5) and U(1)$_X$. 

In the rest of this subsection we will be interested in the study of the physical Higgs doublet, thus we will use $A$ to denote the SO(5) components only. ${\cal L}_{\rm mix}$ and ${\cal L}_1$ contain the following kinetic terms for the scalars:
\begin{equation}\label{kinetic-ngb}
{\cal L}\supset \frac{f_A^2}{4}\sqrt{-G_0}G^0_{\mu\nu}(D^\mu U_A)^\dagger D^\nu U_A +\frac{f_1^2}{4}\sqrt{-G_1}G^1_{\mu\nu}\sum_{\hat a} d^{\hat a\mu}d^{\hat a\nu} \ .
\end{equation}
where the covariant derivative $D_\mu U_A$ and the symbol $d_{\mu}$ are defined by:
\begin{align}
& D_\mu U_A=\partial_\mu U_A-iA_\mu^0 U_A+iU_A A_\mu^1 \ ,
\\
& i U_1^\dagger D_\mu U_1=d_{\mu}+e_{\mu} \ , \qquad d_{\mu}=d_{\mu}^{\hat a} T^{\hat a} \ , \qquad e_{\mu}=e_{\mu}^a T^a \ .
\end{align}

The terms of Eq.~(\ref{kinetic-ngb}) mix the NGB fields $\Pi_A$ and $\Pi_1$ with the gauge fields $A_\mu^0$ and $A_\mu^1$, leading to:
\begin{equation}\label{kinetic-ngb2}
{\cal L}\supset \frac{f_A}{\sqrt{2}}\sum_{r=a,\hat a}(A_\mu^{0r}-A_\mu^{1r})\partial^\mu\Pi^r_A +
 \frac{f_1}{\sqrt{2}}\sum_{\hat a}A_\mu^{1\hat a}\partial^\mu\Pi^{\hat a}_1 \ .
\end{equation}
As usual, this mixing can be cancelled by working in the unitary gauge. Taking into account that the gauge fields of G$_0$/H$_0$ are not dynamical, the unitary gauge corresponds to $\Pi_A^a=0$, $\Pi_A^{\hat a}=\Pi^{\hat a}f_h/f_A$ and $\Pi_1^{\hat a}=\Pi^{\hat a}f_h/f_1$, with:
\begin{equation}
\frac{1}{f_h^2}=\frac{1}{f_A^2}+\frac{1}{f_1^2} \ . 
\end{equation}
In the unitary gauge there is only one scalar multiplet: $\Pi=\Pi^{\hat a}T^{\hat a}$, that can be identified as the Higgs field and has a decay constant $f_h$. ${\cal L}_{\rm mix}$ explicitly breaks the symmetry and induces a potential at 1-loop for the Higgs. This potential can trigger EWSB and lead to a realistic model if $\xi\equiv\sin^2(v/f_h)\sim0.1$~\cite{Agashe:2004rs}. The details of the potential will not be needed for the study of this paper.

In sec.~\ref{sec-mass-basis} we will describe the spectrum of spin-one states.

\subsection{Fermion sector}\label{sec-fermion-sector}
The mixing term for the fermions and the Yukawa interactions on site-1 can be written schematically as:
\begin{equation}\label{eq-ferm1}
{\cal L}\supset\Delta_\psi\bar\psi_0U_A\psi_1+f_1\sum_Ry_RP_R(\bar\psi_1U_1)P_R(U_1^\dagger\psi_1)+{\rm h.c.} \ .
\end{equation}
$P_R$ are projectors that project a given representation of G$_1$ into its components under the subgroup H$_1$. $y_R$ are dimensionless Yukawa couplings, $y_R\sim g_1$. For the case of SO(5)/SO(4) a large set of possibilities have been described in Refs.~\cite{Montull:2013mla,Carena:2014ria}. The different representations for the fermions have an impact on the Higgs potential as well as on the phenomenology, for example on the $Z$-couplings and EW precision tests~\cite{Agashe:2006at,Panico:2012uw,Carena:2014ria}. However the phenomenology that we will study is rather independent of this details, as long as one assumes that $X^1$ is not heavy enough to decay to pairs of composite fermions. In the following, to simplify our analysis, we will assume this to be the case. In sec.~\ref{sec:loop} we will study the 1-loop corrections to the coupling between the massive graviton and gluons, only in this case we will need to specify the representations.

\subsection{Mass basis and graviton interactions}\label{sec-mass-basis}
To study the phenomenology of the graviton at LHC it is convenient to study its interactions in the mass basis. In this section we show the rotations that allow to diagonalize the mixing and compute graviton couplings in that basis. We will not consider the mixing effects arising from EWSB, that give corrections of order $g_0v/m_1$.

Let us show first how the mixing Lagrangian generates masses for several fields. We start with the gravity sector, in the unitary gauge~\cite{ArkaniHamed:2002sp}:
\begin{align}\label{grav-mix}
&{\cal L}_{\rm mix}^{\rm grav}=-\frac{f_X^4}{2}\sqrt{-G_0}(K_{\mu\rho}K_{\nu\sigma}-K_{\mu\nu}K_{\rho\sigma})(K^{\mu\rho}K^{\nu\sigma}-K^{\mu\nu}K^{\rho\sigma}) \ , \\
&K_{\mu\rho}=G^0_{\mu\rho}-G^1_{\mu\rho} \ ,
\end{align}
$f_X\sim$TeV. This term breaks the symmetries of general coordinate transformations on both sites to the diagonal subgroup, generating a mass for a linear combination of $X^0$ and $X^1$ and leaving the orthogonal combination massless.

For the gauge fields, from Eq.~(\ref{kinetic-ngb}) in unitary gauge we obtain:
\begin{equation}\label{kinetic-ngb3}
{\cal L}\supset \frac{f_A^2}{4}\sum_{r=a,\hat a}(A_\mu^{0r}-A_\mu^{1r})^2+
 \frac{f_1^2}{4}\sum_{\hat a}(A_\mu^{1\hat a})^2 \ .
\end{equation}
The first term of Eq.~(\ref{kinetic-ngb3}), arising from ${\cal L}_{\rm mix}^{\rm matter}$, breaks G$_0\times$G$_1\to$G$_{0+1}$. It generates a mass for fields of G$_0\times$G$_1/$G$_{0+1}$ and leaves a set of massless fields in G$_{0+1}$. The second term contributes to the mass of $A^{1\hat a}_\mu$.

For the fermions, in this section we consider the simple case where for each SM fermion there is just one composite partner in a full multiplet of G$_1$. As for the gauge sector, we add spurious fermion fields on site-0 to fill full multiplets of the extended symmetry. The mass and mixing terms for the fermions arise from the mass term of ${\cal L}_1$ in Eq.~(\ref{eq-Lj}) and from Eq.~(\ref{eq-ferm1}), before EWSB they lead to: 
\begin{equation}\label{eq-mpsi}
{\cal L}\supset \sum_{\psi}\Delta_\psi \bar\psi^0\psi^1-\sum_{R,\psi^1}m^R_{\psi_1}P_R(\bar\psi^1)P_R(\psi^1) \ .
\end{equation}
The second term of Eq.~(\ref{eq-ferm1}), in the H$_1$-symmetric phase, generates a splitting between the different representations of H$_1$ contained in G$_1$. For that reason there can be different $m^R_{\psi_1}$ for the different multiplets of H$_1$ in the second term of (\ref{eq-mpsi}). We will consider $m^R_{\psi_1}\sim g_\psi f_1$, with $g_\psi\sim g_1$.

To obtain the physical masses one needs canonically normalized kinetic terms, thus we redefine: $A^j_\mu\to g_j A^j_\mu$ and $G^j_{\mu\nu}\to G^j_{\mu\nu}/M_j$. The elementary sector can be decoupled from the composite one by taking the elementary couplings and fermionic mixing to zero. In this limit the states on site-0 are massless and the states on site-1 have masses: $m_{A_1^a}=g_1f_A/\sqrt{2}$, $m_{A_1^{\hat a}}=g_1(f_A^2+f_1^2)^{1/2}/\sqrt{2}$, $m_{X_1}=f_X^2/M_1$ and $m_{\psi_1}$. We are not distinguishing explicitly the couplings $g_1$ and the scales $f_A$ of the different gauge groups, but the reader must take into account that they can differ. Also notice that, in the present effective description of the composite sector, the masses of the different species of resonances are independent of each other. This situation is less restrictive than the simplest realizations in extra dimensions.~\footnote{In extra dimensions the spectrum of gauge and graviton fields are usually fixed by the size of the extra dimension, although they can be distorted, for example by adding kinetic terms on the boundaries.} 

To obtain the mass basis we perform a rotation of the fields on both sites hat have mixing. Using $\Phi_j$ for any of the fields on site-$j$:
\begin{align}
&\Phi=c_\Phi \Phi_0+s_\Phi \Phi_1 \ , \qquad \Phi^*=-s_\Phi \Phi_0+c_\Phi \Phi_1 \ , \qquad t_\Phi=\frac{s_\Phi}{c_\Phi} \ , \label{eq-rotation} \\
& t_A=\frac{g_0}{g_1} \ , \qquad t_\psi=\frac{\Delta}{m_{\psi_1}} \ , \qquad t_X=\frac{M_1}{M_0} \ ,
\end{align}
$\Phi$ are massless fields and $\Phi^*$ are massive, with mass $m_{\Phi^*}=m_{\Phi_1}\sqrt{1+t_\Phi^2}$. The components of $\Phi_1$ that do not mix, or those that mix with spurious fields on site-0, are usually called custodians, they are not rotated and have masses $m_{\Phi_1}$. The variables $s_\Phi$, $c_\Phi$ and $t_\Phi$ are shorthands for the trigonometric functions: $\sin\theta_\Phi$, $\cos\theta_\Phi$ and $\tan\theta_\Phi$, $s_\Phi$ is a measure of the degree of compositeness of the mass eigenstates. We will consider $t_A\sim s_A\sim 0.1- 0.35$. The ratio $t_A$ can be different for the different gauge groups. We will call universal to the case where these quantities are the same for all groups, but we will also consider departures from universality that, as we will show, can have interesting consequences for the phenomenology.

We find it useful to define also an angle for the Higgs:
\begin{equation}
t_H=\frac{f_A}{f_1}, \qquad s_H=\frac{f_h}{f_1}, \qquad c_H=\frac{f_h}{f_A} \ .
\end{equation}
For $f_A=f_1$: $s_H=c_H=1/\sqrt{2}$. Values of $s_H$ very close to zero or one require a hierarchy between $f_A$ and $f_1$.

The gauge bosons and the graviton of the unbroken groups, $A_\mu$ and $X_{\mu\nu}$, are massless. Their couplings are: $g^{-2}=g_0^{-2}+g_1^{-2}$ and $M^{-2}=M_0^{-2}+M_1^{-2}$.

After EWSB there are corrections to the previous description. The most important ones are the masses for the SM fermions and EW bosons. The masses of these fermions can be approximated by: $m_\psi\sim s_{\psi_L}s_{\psi_R}vg_1$, with $s_{\psi_{L,R}}$ the mixing angle of the corresponding chiralities. The hierarchy of fermion masses can be obtained by considering hierarchically small mixing angles. For the top: $s_q,s_t\sim 0.5 - 1$. In the following we will consider that the Left- and Right-handed mixing of all the other fermions are very small, leading to almost elementary SM fermions. As a consequence they will not play an important role in our analysis and we will not consider them. A possible exception can be the bottom quark, with Left-handed mixing equal to that of the top. In some models, as in MCHM$_5$ of Ref.~\cite{Contino:2006qr}, the Left-handed doublet mixes with two composite states, one mixing leading to the top mass, and another one, $s_{q'}$ leading to the bottom mass. In this case we assume $s_{q'}$ very small, and we take into account the effect of the Right-handed mixing $s_b$, that can be sizable~\cite{Andres:2015oqa}.

We describe now the graviton interactions after the elementary/composite rotations have been done, we will neglect the new mixing arising form EWSB. We write the interactions linear in the massive graviton as:
\begin{equation}\label{eq:Lgrav}
{\cal L}\supset \sum_\Phi \tilde C_\Phi X^*_{\mu\nu}T^{\mu\nu}(\Phi) \ .
\end{equation}
In this case $\Phi$ includes also the Higgs field. The different terms of the energy-momentum tensor are similar to those defined in Eq.~(\ref{eq-energymomentum-tensor}). The contribution to $T^{\mu\nu}$ from the Higgs, for processes involving two scalar particles, can be taken equal to the contribution from the SM Higgs, see for example Ref.~\cite{Falkowski:2016glr}. The couplings are given by:
\begin{align}
& \tilde C_{A}=-\frac{s_A^2c_X}{M_1}+\frac{c_A^2s_X}{M_0} \ ,
\qquad \tilde C_{\psi}=\frac{s_\psi^2c_X}{M_1}-\frac{c_\psi^2s_X}{M_0} \ , \label{eq-X-Af}\\
& \tilde C_{A^*}=-\frac{c_A^2c_X}{M_1}+\frac{s_A^2s_X}{M_0}  \ ,
\qquad \tilde C_{\psi^*}=\frac{c_\psi^2c_X}{M_1}-\frac{s_\psi^2s_X}{M_0}  \ , \\
& \tilde C_{A-A^*}=2s_A c_A\left(\frac{c_X}{M_1}-\frac{s_X}{M_0}\right)  \ ,
\qquad \tilde C_{\psi-\psi^*}=-2s_\psi c_\psi\left(\frac{c_X}{M_1}-\frac{s_X}{M_0}\right) \ , \label{eq:XtoKKandSM}\\
& \tilde C_H=\frac{s_H^2c_X}{M_1}-\frac{c_H^2s_X}{M_0} \ , \label{eq-X-H}
\end{align}
where $C_\Phi$ and  $C_\Phi^*$ are couplings involving the same field, and $C_{\Phi-\Phi^*}$ involves a light and a heavy field, besides the graviton.

After EWSB one has to rotate to the photon-$Z$ basis for the neutral spin-one states. This rotation induces an interaction with $\gamma$ and $Z$:
\begin{align}
&\tilde C_\gamma=\tilde C_W \sin^2\theta_w+\tilde C_B\cos^2\theta_w \ ,
&\tilde C_Z=\tilde C_W \cos^2\theta_w+\tilde C_B\sin^2\theta_w
\\
&\tilde C_{Z\gamma}=\sin\theta_w\cos\theta_w(\tilde C_W-\tilde C_B) \, , &\label{zgammacoeff}
\end{align}
where $\theta_w$ is the Weinberg angle. For universal couplings $\tilde C_{Z\gamma}$ vanishes.

For very small mixing, $s_\Phi\to 0$, the massless states interact with the massive graviton with couplings $s_X/M_0$. However, taking into account that $M_1\ll M_0$, in general the first term  dominates: $\tilde C\simeq\pm s_\Phi^2/M_1$, leading to a coupling modulated by the degree of compositeness of the state coupled to the graviton, as well as by $M_1$. We find it convenient to define a dimensionless coupling:
\begin{equation}
C_\Phi=\tilde C_\Phi M_1 \ . \label{coeff}
\end{equation}
We will use this dimensionless coupling to present our results in the phenomenological analysis of the next sections.

\section{Phenomenology}\label{sec:pheno}
The main point in this section is to find out the most sensitive graviton decay channel through which it could be resonantly detected at the LHC.  The outcome to this question depends --at least-- on the graviton mass, its coupling scale $M_1$ and the mixing parameters, which determine the graviton production and branching ratios.  Along the next paragraphs we address this question and we also understand some general qualitative patterns.

Within the theoretical framework described above we can study the phenomenology of this scenario by parametrizing the graviton production cross section and its branching ratios through the free variables of the model.  Under the universal couplings assumption these variables would be $s_A,\ s_{q},\ s_{t},\ s_{b},\ C_H$ and $M_1$, whereas for the non-universal case one should disaggregate $s_A \to s_G,\ s_W$ and $s_B$ for the three gauge groups.  Unless explicitly stated, we refer to the universal case.  In the following paragraphs we study tree-level phenomenology and leave one-loop effects for next section.

At tree level, we can easily parametrize the graviton production cross section by computing it at some given energy, mass and coupling normalized to one, and then re-scaling with the square of the graviton coupling to gluons.  For instance, using MadGraph \cite{Alwall:2014hca} with PDF NN23LO1 we have that for physical massive graviton $X^*$ with mass $m_{X^*}=1$ TeV and LHC energy 8 and 13 TeV:
\begin{eqnarray}
\sigma(pp\to X^*) &=& \left\{ 
\begin{array}{cc} 
\left(\frac{3\text{ TeV}}{M_1}\right)^2 (0.004\, s_b^4 + 5.6 \, s_A^4)  \, \mbox{pb} & 8 \mbox{TeV,} \\
\left(\frac{3\text{ TeV}}{M_1}\right)^2 (0.023\, s_b^4 + 30.2 \, s_A^4  )\, \mbox{pb} & 13 \mbox{TeV} 
\end{array} 
\right.
\label{xsection}
\end{eqnarray}
where we have assumed a QCD $k$-factor $k=1.6$ \cite{Falkowski:2016glr,Das:2016pbk}.  Although in general one can safely neglect the $b\bar b \to X^*$ process, this channel is the only production mechanism consi\-dered when looking for $b\bar b$ resonances since these specific search strategies assume this produc\-tion process.

The formulae for the width of the graviton to the different particles can be found elsewhere \cite{Falkowski:2016glr}, however we quote here the relevant ones for the discussion that follows,~\footnote{Although we have considered graviton masses up to 3 TeV in this work, and in some cases there could be lighter resonances, for simplicity we have not included the possibility of $X^*$ decaying to composite resonances.}
\begin{eqnarray}
\Gamma(X^* \to f\bar f) \!\!\!&=&\!\!\! \frac{N_c m_{X^*}^3}{320 \pi M_1^2} (1-4r_f)^{3/2} \left( (|C_{f_L}|^2 + |C_{f_R}|^2) \left( 1-\frac{2 r_f}{3} \right) + Re(C_{f_L} C_{f_R}^*)  \frac{20 r_f}{3} \right)\!, \hspace*{0.7cm}
 \label{ff} \\ 
\Gamma(X^* \to ZZ ) \!\!\!&=&\!\!\!  \frac{ m_{X^*}^3}{80 \pi M_1^2} (1-4r_Z)^{1/2} \bigg( |C_{Z}|^2 + \frac{|C_H|^2}{12} + \frac{r_Z}{3} \Big( 3|C_H|^2 - 20 Re(C_H C_{Z}^*) - 9 |C_{Z}|^2 \Big) \nonumber \\
&& + \frac{2 r_Z^2}{3} \Big(  7|C_H|^2 +10  Re(C_H C_{Z}^*) + 9 |C_{Z}|^2  \Big) \bigg), \label{zz}\\ 
\Gamma(X^* \to \gamma\gamma) \!\!\!&=&\!\!\! \frac{|C_\gamma|^2 m_{X^*}^3}{80 \pi M_1^2}, \label{aa} \\ 
\Gamma(X^* \to HH) \!\!\!&=&\!\!\! \frac{|C_H|^2 m_{X^*}^3}{960 \pi M_1^2}  (1-4r_H)^{5/2} , \label{hh}  
\end{eqnarray}
whereas $\Gamma(X^* \to WW )=2\Gamma(X^* \to ZZ )$ replacing $m_Z\to m_W$ and $C_Z \to C_W$.
Here $r_i = m_i^2/m_{X^*}^2$.
We recall that for universal mixing of all gauge groups is valid
\begin{equation}
C_\gamma=C_Z = C_W \simeq - s_A^2,
\end{equation}
where $s_A$ is the mixing angle for the gauge boson defined in Eq.~(\ref{eq-rotation}).
In particular, under this assumption the $X^* \to Z\gamma$ decay is not allowed at tree-level. However this is only a simplified picture, and the decay $X^* \to Z\gamma$ can be open if the mixing differs for the different groups of the EW sector.  In this case we can write 
\begin{eqnarray}
\Gamma(X^* \to Z\gamma) \!\!\!&=& \!\!\! \frac{m_{X^*}^3}{40 \pi M_1^2} |C_{Z\gamma}|^2  (1 - r_Z)^3 (1 + \frac{r_Z}{2} + \frac{r_Z^2}{6} ) \, , \label{za2}
\end{eqnarray}
where $C_{Z\gamma}$ is defined in Eqs.~(\ref{zgammacoeff}) and (\ref{coeff}). With this information at hand, there are some general features of the model that can already be discussed at this point.

Observe that in this model all graviton couplings to SM particles have an upper bound of $1/M_1$. We will take $M_1=3$ TeV, and we will discuss briefly the dependence on this variable. 

As discussed below, for this value of $M_1$ the fermion couplings modulated by the mixing angles $s_{q},\ s_{t}$ and $s_{b}$ will not have a dominant role in determining the most sensitive channel unless other couplings are very small.  This is because experimental limits on fermion resonance searches are not saturated for this value of $M_1$ not even for maximal mixing. Observe, however, that the branching ratios to fermions may be dominant. This is numerically verified below.

\begin{figure}[h!]
\begin{minipage}[c]{8cm}
\includegraphics[width=8cm]{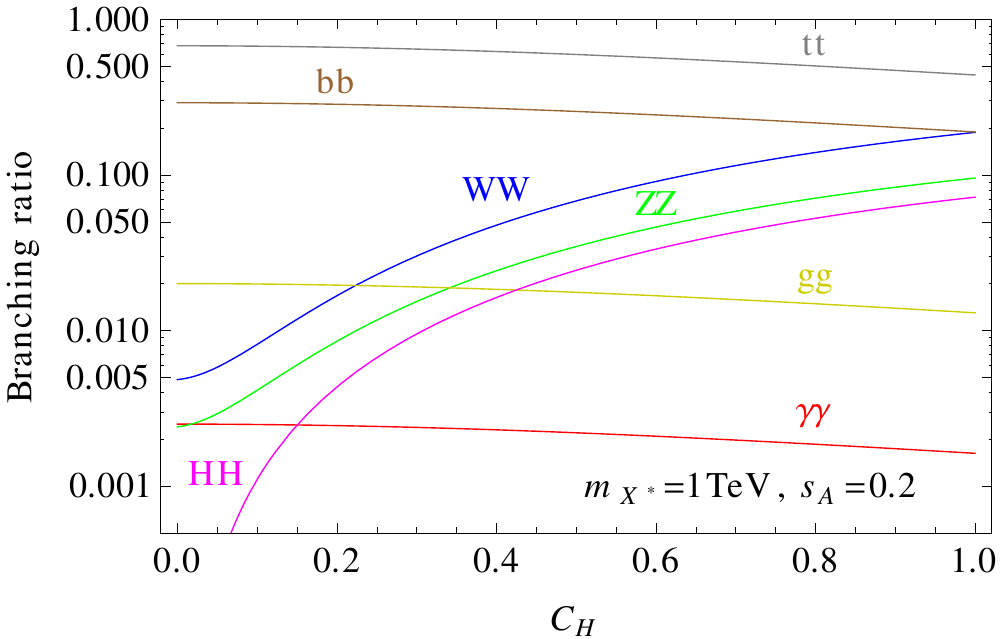}
\end{minipage}
\begin{minipage}[c]{8cm}
\includegraphics[width=8cm]{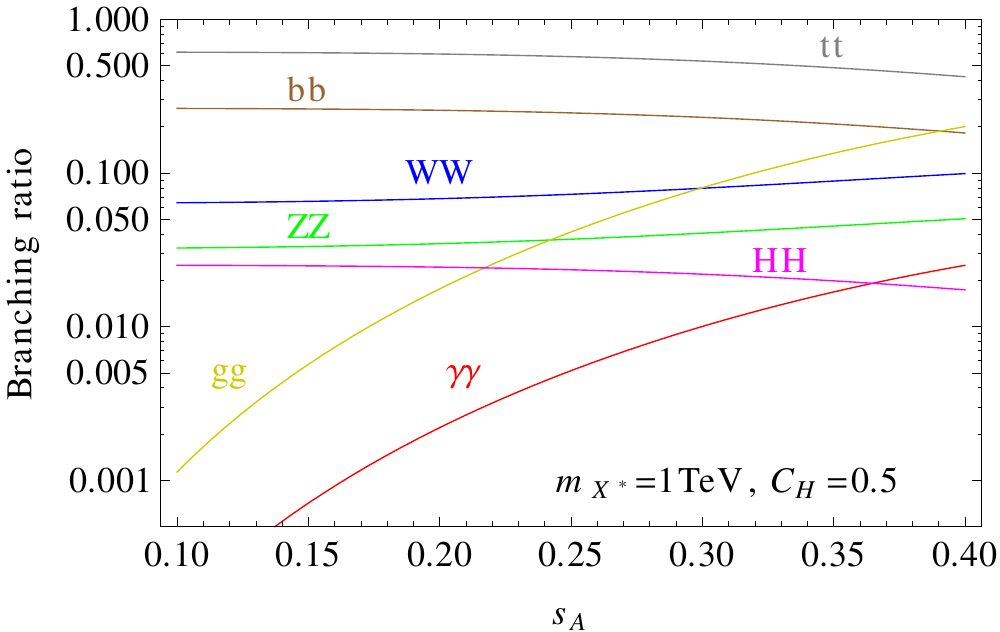}
\end{minipage}
\\
\begin{minipage}[c]{8cm}
\vspace*{0.5cm}
\includegraphics[width=8cm]{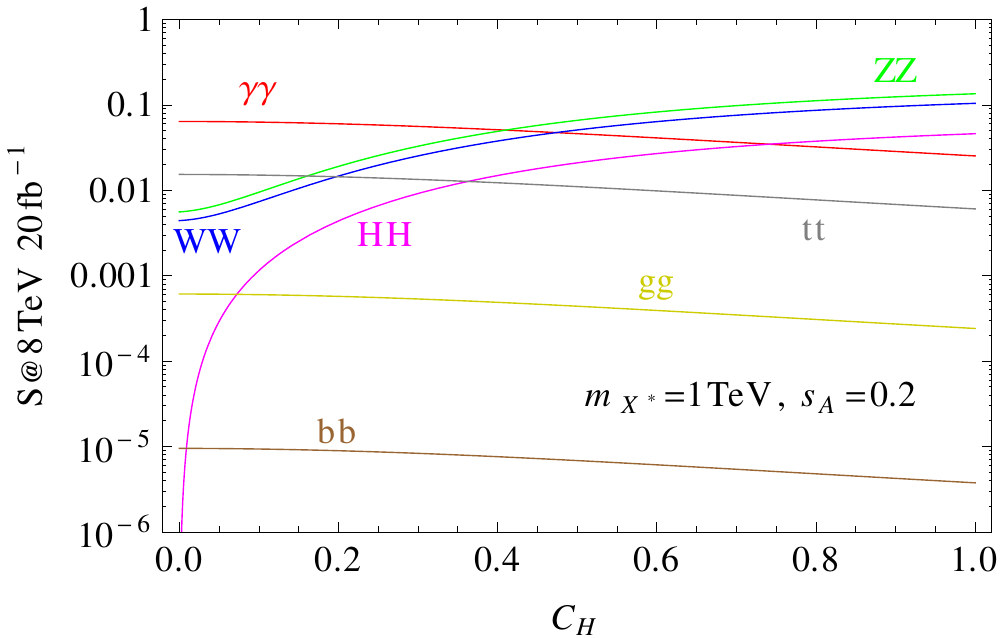}
\end{minipage}
\begin{minipage}[c]{8cm}
\vspace*{0.5cm}
\includegraphics[width=8cm]{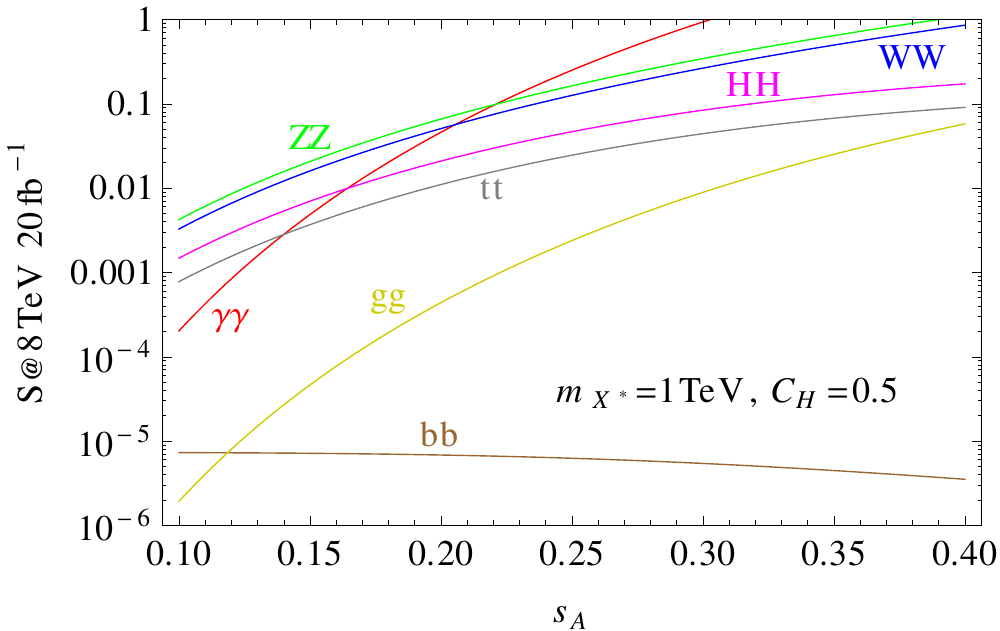}
\end{minipage}
\caption{
\small
Upper panel: Graviton branching ratios as a function of the two more relevant parameters, $C_H$ and $s_A$, within the universal coupling scenario. Lower panel: Sensitivity of graviton decay channels at 8 TeV and 20 fb$^{-1}$ expressed in terms of ${\cal S}$  as a function of $C_H$ and $s_A$ within the universal scenario. All plots were generated with the following setting of graviton couplings to fermions: $s_{q}=0.7,\ s_{t}=0.8$ and $s_{b}=0.3$; the patterns depicted in the figure are a general feature of a massive graviton independently of these specific values though.
}
\label{brs}
\end{figure}

In light of the above discussion, it is instructive to study the graviton branching ratios as a function of the relevant variables $s_A$ and $C_H$. Notice that $M_1$ modifies the production cross section, but not the branching ratios.  Similarly, $m_{X^*}$ also affects the production cross section, and only slightly the branching ratios through the $r_i$ parameters.  In the upper panel of Fig.~\ref{brs} we plot the branching ratio behavior as a function of the variables $s_A$ and $C_H$ for the case of universal couplings. For non-universal couplings a new decay channel is open: $X^*\to Z\gamma$. To leading order in $r_Z$: $\Gamma(X^*\to Z\gamma)=\Gamma(X^*\to\gamma\gamma)|C_{Z\gamma}/C_{\gamma}|^2/2$.

There are two main features which can be understood from the branching ratios plots, upper panel of Fig.~\ref{brs}.  The first one is that the model variable $C_H$ does not only affect $X^* \to hh$, but also and to practically the same extent $X^* \to ZZ/WW$ due to their longitudinal polarizations.  In the left panel of the figure we can see that for large $C_H$ there is an important enhancement to $hh,\ ZZ$ and $WW$, whereas other branching ratios decrease.  Due to different experimental sensitivity on these channels, we will see below that this favors mostly the $ZZ$ channel, and also the $WW$ channel at large energies. The second point is seen in the right upper plot of Fig.~\ref{brs}, where the dependence on $s_A$ affects mostly the $gg$ and $\gamma\gamma$ channels. In fact, the $WW$ and $ZZ$ channels, which also depend on this variable, are only slightly affected because they also have an important contribution from $C_H$. Therefore, increasing $s_A$ determines an increment in the graviton production cross section through the $gg\to X^*$ process (see Eq.~(\ref{xsection})) and also an enhancement of $gg$ and $\gamma\gamma$ decay channels.  Again, due to experimental sensitivity, we will find below that this favors mainly the $\gamma \gamma$ channel.

\subsection{Comparing different channels using phenomenological natural units}

In addition to the previous discussion on the graviton branching ratios, the different experimental sensitivity of the different decay channels, as well as its dependence with the graviton mass, will play a key role in determining which is the most sensitive channel to find a resonant signal.  It is then natural to compare all decay channels in terms of their experimental sensitivity for a given graviton mass.

This addressing of the problem may lead to two different paths. One is to design search strategies for all channels as a function of the graviton mass and the LHC energy and luminosity, and then compare which channel would be the most sensitive.  Alternatively, we can study the available experimental searches in the different channels and take from them the experimental limits for a given energy and luminosity. Since this last path  is based on real performed searches, we expect it to provide additional experimental information which would be difficult to include in the former option. However, some difficulties may rise due to searches performed with different luminosities.

In this work we will take the second path and compare the strength of the signal in each channel in units of the measured experimental limits in each channel. That is, the strength ${\cal S}$ for a given channel, graviton mass and center of mass energy is defined as the ratio of the predicted graviton production cross section times branching ratio times acceptance ($\sigma_{\mbox{\scriptsize pred}}$) to the corresponding experimental limit at the 95\% CL ($\sigma_{\mbox{\scriptsize lim}}$) in that channel for that mass at a certain luminosity, namely, 

\begin{equation}
{\cal S} = \frac{\sigma_{\mbox{\scriptsize pred}}}{\sigma_{\mbox{\scriptsize lim}}} \, . \label{S-de-Sequi}
\end{equation}

The meaning of the strength is straightforward.  If for any channel ${\cal S}>1$ at a given point in parameter space, then that point is experimentally excluded.  If ${\cal S}<1$ for all channels, then the point is not excluded and the channel with larger ${\cal S}$ at equal conditions of luminosity and energy is the most sensitive channel. Assuming that experimental limits in different channels have a similar scaling with luminosity, then the channel with larger ${\cal S}$ would be the first one to observe or exclude the postulated NP.

We illustrate in Fig.~\ref{brs} the phenomenological importance of the information contained in ${\cal S}$. Upper and lower panels show that graviton decay channels with dominant branching ratios become suppressed in terms of ${\cal S}$ and vice versa. For instance, comparing the left plots of Fig.~\ref{brs}, we can see that $t\bar t$, the channel with the largest branching ratio, is exceeded  by $\gamma \gamma$ in a ${\cal S}$ plot although BR($\gamma \gamma$) is significantly smaller than BR($t\bar t$). Therefore, ${\cal S}$ quantifies the compromise between theoretical expected relevance and experimental cleanliness in determining the relative phenomenological impact of different decay channels. We will discuss in detail the implementation of ${\cal S}$ to our analysis in secs.~\ref{sec:8} and \ref{sec:13}.

It is worth stressing at this point that these strength units have encoded inside a diversity of experimental aspects and, in particular, many of them suffer modifications as a function of the mass of the particle that is sought. Moreover, working in these units includes important changes due to modifications in the search strategy of a given channel as the expected momentum of the reconstructed particles increase.  For instance, a search for dibosons at low $p_T$ is mainly performed in the leptonic channel, whereas at large $p_T$ is better performed in the hadronic channel through fat jet techniques. The use of specific final states in an experimental search may lead to larger branching ratios, for instance in $Z$ bosons, the branching ratio goes from 6\% in the case of $Z$ decaying to electrons and muons to 70\% for hadronic decays. Therefore, the $ZZ$ channel suffers an important increase in sensitivity relatively to other channels. Similar drastic transitions occur also in $WW$ and $t\bar t$.  Also other minor changes occur in all other channels.  In addition to these alterations in the expected signal, all channels suffer a variety of changes in their respective backgrounds as energy changes, which yields considerable modifications in the final relevant variable: the sensitivity.  Summarizing, these units are simple to implement but not trivial to understand since they contain many important information encoded inside which should be taken into account in order to achieve a better use of their capabilities.

Although for LHC at 8 TeV experimental searches exist for the final states corresponding to all the decay channels for the same luminosity (20 fb$^{-1}$), this is not the case for LHC at 13 TeV.~\footnote{Since ATLAS and CMS have not yet reported dedicated searches for gravitons in some of the decay channels under consideration here and no qualitative difference in our results is expected, we have extracted the experimental limits in these cases from dedicated searches for scalar or vector particles.}  However, since searches at 13 TeV are performed at luminosities within fairly the same order of magnitude, we will assume a statistic uncertainty regime and re-scale the experimental sensitivity with the square root of the ratio of luminosities.  If the maximum allowed cross section of a signal at a given luminosity $L_i$ is $\sigma_s^{(i)}$, then under this assumption is valid 
\begin{equation}
\sigma_s^{(2)} = \frac{\sigma_s^{(1)}}{\sqrt{L_2/L_1}}, \label{lum}
\end{equation}
for each channel. Since the strength is inversely proportional to the maximum allowed cross section, it is easy to obtain that ${\cal S} \propto \sqrt L$.  Therefore, given a point with ${\cal S}<1$ at a given luminosity, an increase in luminosity by a factor $1/{\cal S}^2$ is required to discard/observe it.

In Table \ref{sensitivities} we show the collected sensitivities in different channels, for different energy and luminosities and for three reference graviton masses.  In all cases we have taken the expected limit instead of the observed one, to avoid what could be statistical fluctuations.  This collection of limits does not pretend to be exhaustive, but rather a fair sample of the state-of-the-art.

\begin{table}
\begin{center}
\begin{tabular}{ |l|l|l|l|l| }
\hline
 &\multicolumn{2}{ |c| }{$E = 8$ TeV \& $L=20$ fb$^{-1}$}&\multicolumn{2}{ |c| }{ $E = 13$ TeV \& $L=13.3$ fb$^{-1}$} \\
\hline
Decay channel & $m_{X^*}=0.5$ TeV & $m_{X^*}=1$ TeV &$m_{X^*}=1$ TeV &$m_{X^*}=3$ TeV \\
\hline
$ZZ$ & 0.046 pb \cite{Aad:2015kna} &  0.011 pb \cite{Aad:2015kna} & 0.055 pb \cite{Aaboud:2016okv} & 0.0032 pb \cite{Aaboud:2016okv} \\ \hline
$WW$ & 0.21 pb \cite{Aad:2015agg} &  0.028 pb \cite{Aad:2015agg} & 0.055 pb \cite{Aaboud:2016okv} & 0.0032 pb \cite{Aaboud:2016okv} \\ \hline
$Z\gamma$ & 0.0063 pb \cite{Aad:2014fha} &  0.0027 pb \cite{Aad:2014fha} &  0.0096 pb \cite{ATLAS:2016lri} & 0.0047$^\dagger$ \cite{ATLAS:2016lri} \\ \hline
$\gamma \gamma $ & 0.0042 pb \cite{Aad:2015mna} &  0.001 pb \cite{Aad:2015mna} & 0.0028 pb \cite{Aaboud:2016tru} & 0.00058 pb \cite{Aaboud:2016tru} \\ \hline
$hh$ & 0.24 pb \cite{ATLAS:2014rxa} &  0.024 pb \cite{ATLAS:2014rxa} & 0.043 pb \cite{ATLAS:2016ixk} & 0.0092 pb \cite{ATLAS:2016ixk} \\ \hline
$jj$ & 11.09 pb \cite{CMS:2015neg,Aad:2014aqa} &  0.839 pb \cite{CMS:2015neg,Aad:2014aqa} & n/a  & 0.090 pb \cite{ATLAS:2016lvi} \\ \hline
$t\bar t$ & 2.05 pb \cite{Chatrchyan:2013lca} &  0.376 pb \cite{Chatrchyan:2013lca} & 0.668 pb \cite{CMS:2016zte} & 0.0252 pb \cite{CMS:2016zte} \\ \hline
$b \bar b$ & 1.71 pb \cite{Khachatryan:2015tra} &  0.975 pb \cite{Khachatryan:2015tra} &  n/a
 & 2.45 pb \cite{ATLAS:2016gvq} \\ \hline
\end{tabular}
\end{center}
\caption{\small
Experimental limits to a graviton resonant signal for different decay channels.  At 13 TeV the $ZZ$ and $WW$ channel sensitivities are computed for a heavy resonance and therefore both channels are merged together in one unique fat jet search. Also at 13 TeV the luminosity has been unified to 13.3 fb$^{-1}$ as explained in text.  $^\dagger$ This limit has been extrapolated.}
\label{sensitivities}
\end{table}

\subsection{LHC at 8 TeV}\label{sec:8}

The aim of this section is to analyze the sensitivity of the graviton decay channels discussed previously with data collected from LHC at 8 TeV with a luminosity of 20 fb$^{-1}$.  With this purpose, we quantify their sensitivity through the strength ${\cal S}$ defined in Eq.~(\ref{S-de-Sequi}) and find out the most sensitive channels within the allowed parameter space of the model. We consider two different scenarios to perform this analysis: a graviton with universal or non-universal couplings.

\subsubsection{Universal couplings}\label{uni-8}

We will see in this section that the parameters $s_A$ and $C_H$ control the degree of sensitivity of the different graviton decay channels whereas, as mentioned previously, $s_{q},\ s_{t}$ and $s_{b}$ just have a minor impact over almost all the parameter space examined. The gravity scale $M_1$ is also a relevant parameter in this analysis since it directly affects the production cross sections, which decrease as $M_1$ increases. However, since $M_1$ does not modify the branching ratios, it plays the role of a global normalization factor which we set to a conservative value of $M_1=3$ TeV. 

\begin{figure}[h!]
\begin{minipage}[c]{8cm}
\includegraphics[width=8cm]{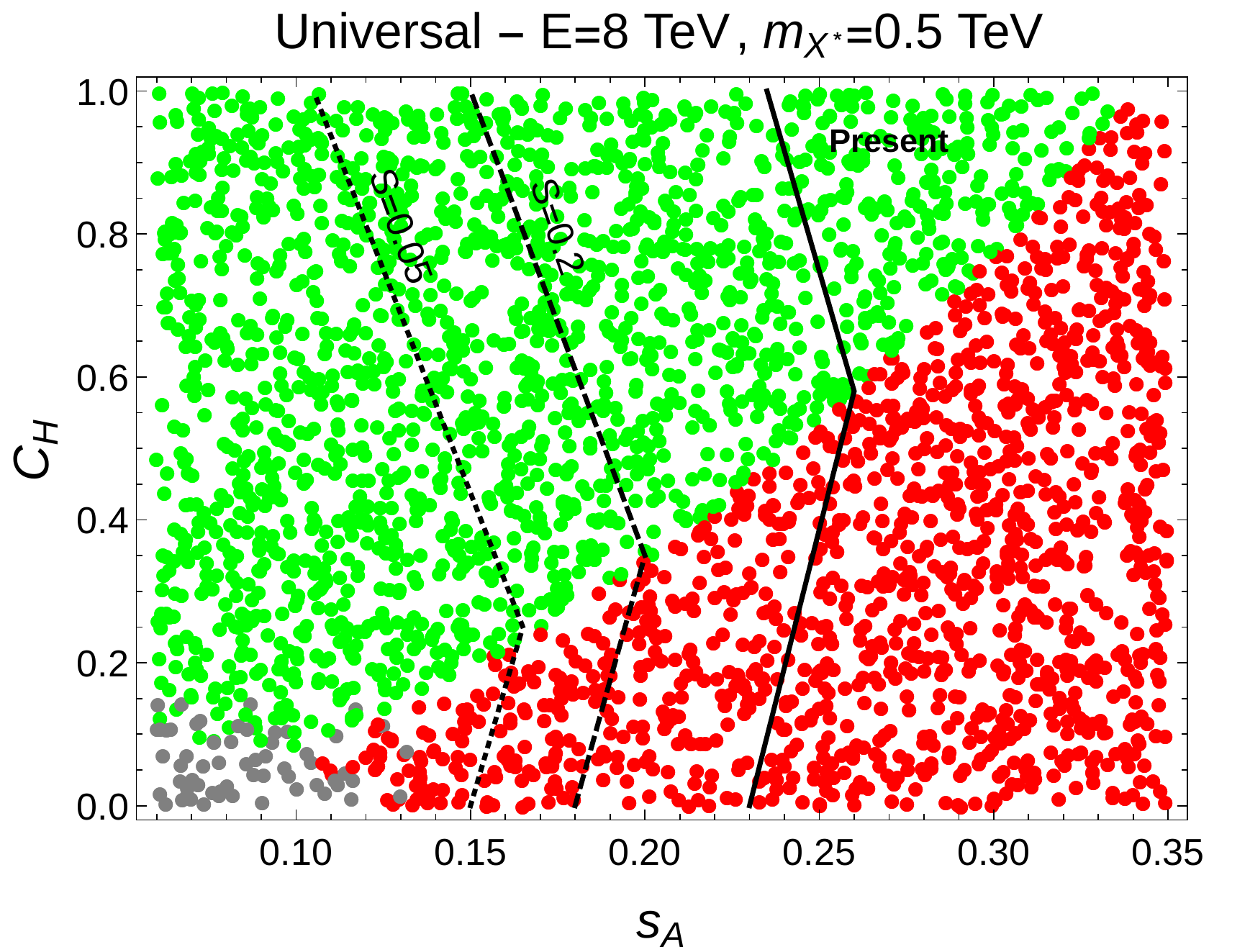}
\end{minipage}
\begin{minipage}[c]{8cm}
\includegraphics[width=8cm]{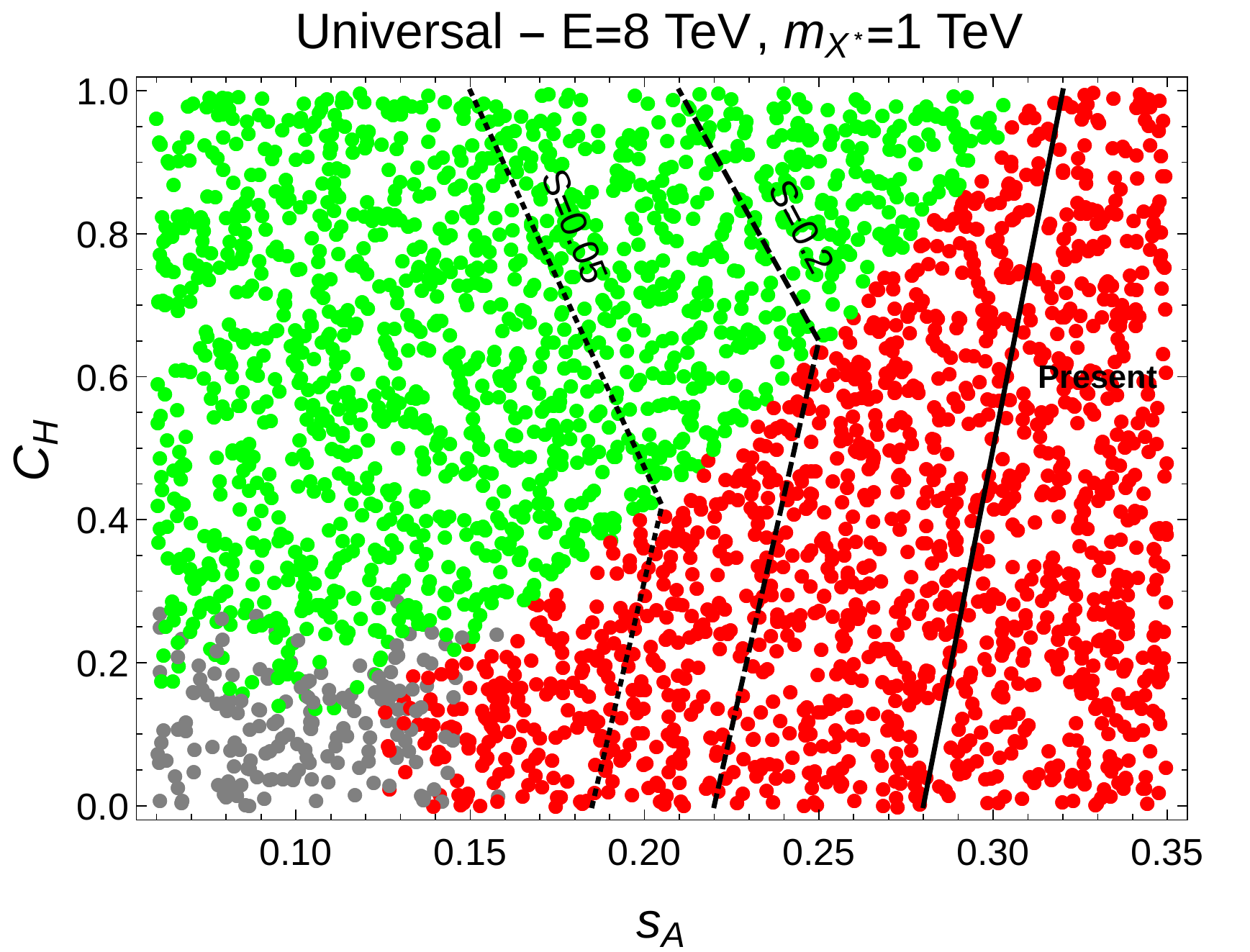}
\end{minipage}
\\
\begin{minipage}[c]{8cm}
\vspace*{0.5cm}
\includegraphics[width=8cm]{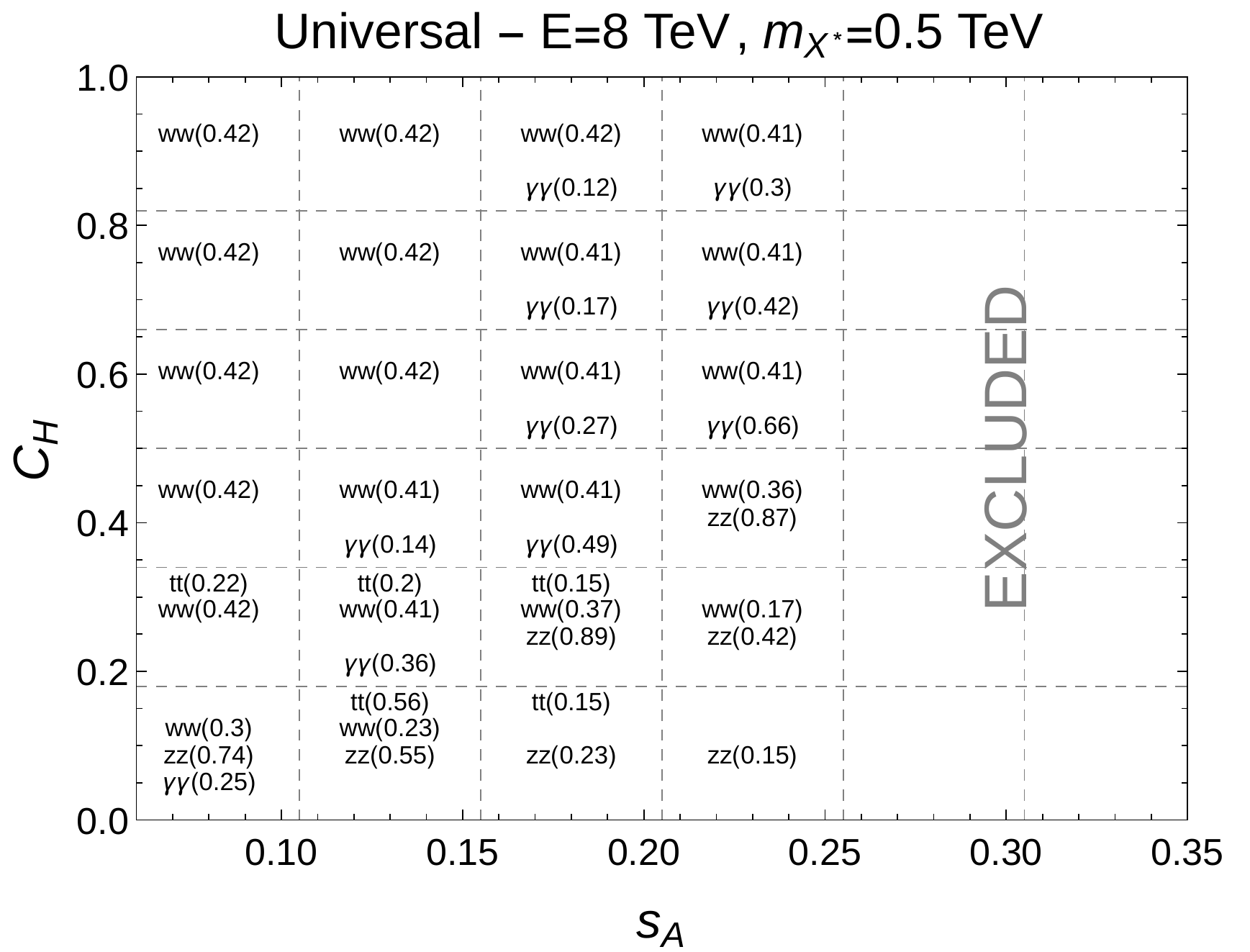}
\end{minipage}
\begin{minipage}[c]{8cm}
\vspace*{0.5cm}
\includegraphics[width=8cm]{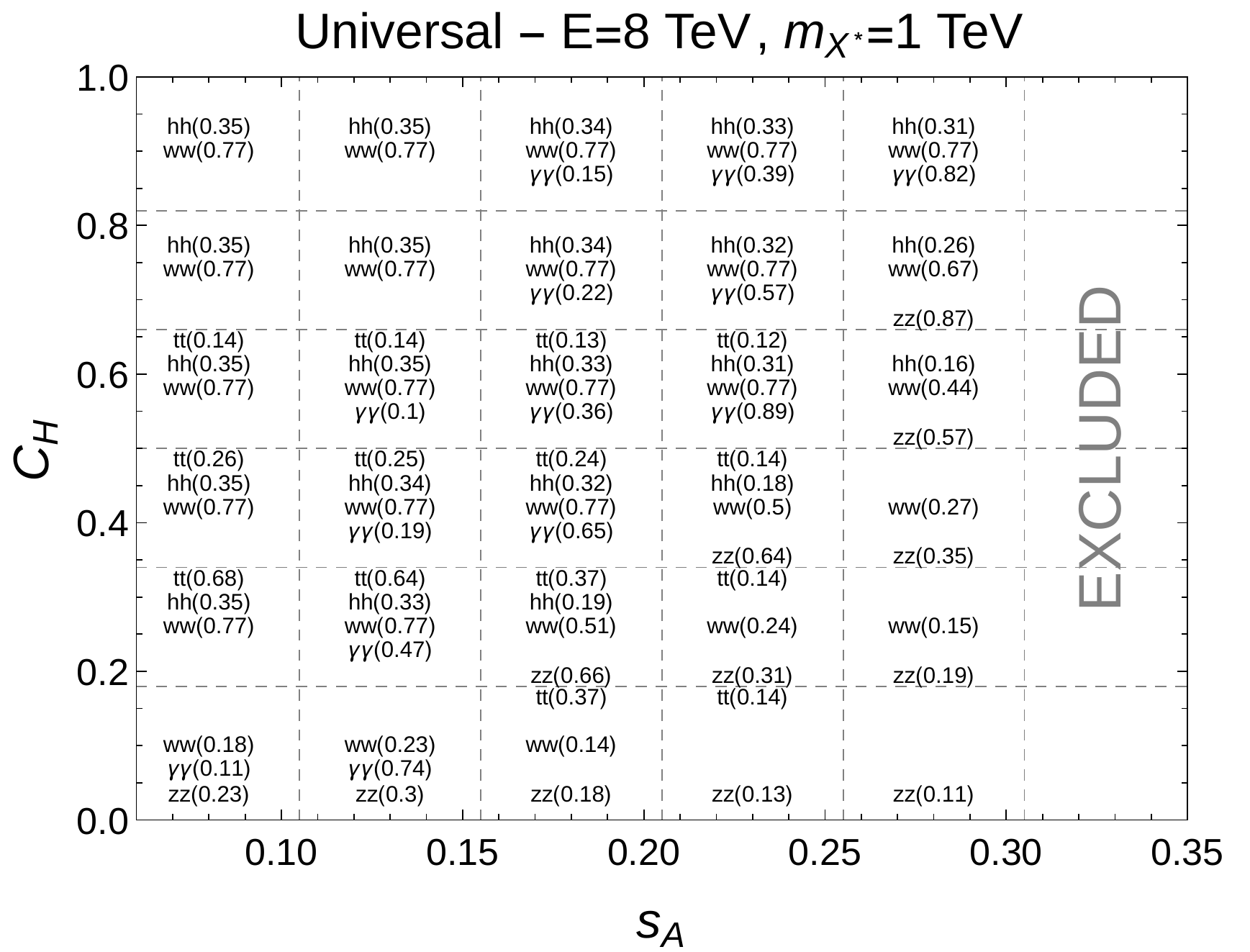}
\end{minipage}
\caption{
\small
Upper panels: The values of $C_H$ and $s_A$ for the points corresponding to the most sensitive graviton decays channels for LHC at 8 TeV within the scanned parameter space (defined in the text), for $m_{X^*}=0.5$ TeV and $m_{X^*}=1\text{ TeV}$, and according to the following color coding: green, red and gray represent $ZZ$, $\gamma \gamma$ and $t\bar t$ channels, respectively. Solid lines define regions excluded by present data. Dotted and dash lines correspond to strength values ${\cal S}=0.05$ and ${\cal S}=0.2$, respectively. Lower panels: Next to most sensitive graviton decay channels for LHC at 8 TeV, for $m_{X^*}=0.5\text{ TeV}$ and $m_{X^*}=1\text{ TeV}$. The numbers in parentheses next to a certain channel indicate its relative sensitivity (${\cal S}_{\mbox{\scriptsize nMS}}/{\cal S}_{\mbox{\scriptsize MS}}$) with respect to the most sensitive one for the central point of each rectangle of the grid.
}
\label{8tev-universal-leading}
\end{figure}

In view of these considerations, we generate scatter plots in the $s_A$-$C_H$ plane randomly scanning over all variable parameters ($s_A \in (0.06,0.35),\ s_{q}\in (0.5,0.95),\ s_{t}\in (0.5,0.95),\ s_{b}\in (0.1,0.3)$ and $C_H\in (0,1)$) for two representative values of graviton masses, $m_{X^*}=0.5$ TeV and $m_{X^*}=1$ TeV. Each point in these plots indicates which graviton decay channel is the most sensitive one, meaning the channel that has the maximum value of ${\cal S}$ with respect to the others. The results are shown in the upper panels of Fig.~\ref{8tev-universal-leading} and there are several points to be discussed.

First, since each different color stands for a given decay channel (see the figure caption for the color coding), we verify a very small dependence on the not plotted parameters ($s_{q}$, $s_{t}$ and $s_{b}$) reflected in the little overlap of colors. Besides, the black lines in the upper plots define, to the left, regions of points in the parameter space which are allowed (${\cal S} < 1$) by the present bounds at 8 TeV and, to the right, regions which are excluded (${\cal S} > 1$) by the same bounds.~\footnote{From now on, excluded regions are determined by only considering the present limits in the most sensitive channel. These regions may be more constraining if the limits of all the channels were combined together.} In addition, dotted and dash lines are defined for two constant values of ${\cal S}$, ${\cal S}=0.05$ and ${\cal S}=0.2$, respectively. They offer a graphic reference of the distribution of values of ${\cal S}$ over the parameter space and how far they are from ${\cal S}=1$. Now, given that the longitudinal $Z$ polarization has larger couplings as $C_H$ increases, we see that for each graviton mass the $ZZ$ channel becomes the most sensitive in a region where $s_A$ is not large enough to make the $\gamma \gamma$ channel reach a maximum of ${\cal S}$. This behavior can be understood in terms of Eqs.~(\ref{xsection}), (\ref{zz}) and (\ref{aa}). The production cross section is the same for both channels and it rises with $s_A$ (for $M_1$ fixed). On the other hand, BR($ZZ$) and BR($\gamma \gamma$) also increase with $s_A$, but only BR($ZZ$) grows with $C_H$, whereas BR($\gamma \gamma$) does not depend on $C_H$. Besides, since the experimental limit of each channel is fixed for a given graviton mass, $ZZ$ turns out to be the most sensitive channel as $C_H$ increases with $s_A$ kept constant. In the complementary regime, $s_A$ taking larger values with $C_H$ constant, $\gamma \gamma$ starts to exceed $ZZ$ since the resulting increment in the predicted BR($\gamma \gamma$) is enough to approach better than $ZZ$ the corresponding experimental limits.

There are also some distinct features among the plots for $m_{X^*}=0.5$ TeV and $m_{X^*}=1$ TeV. The first observation concerns the excluded regions. The larger excluded region in the case of $m_{X^*}=0.5$ TeV in relation to $m_{X^*}=1$ TeV immediately follows from the stringent experimental limits at lower graviton masses since gravitons of larger masses are less easily produced and then more difficult to be excluded. Moreover, the degree of sensitivity of $ZZ$ compared to $\gamma \gamma$ remains almost the same along the parameter space for the two graviton masses. There is only a small effect in the region of large $s_A$ and $C_H$ where the $\gamma \gamma$ channel becomes more sensitive for $m_{X^*}=1$ TeV in comparison to $m_{X^*}=0.5$ TeV. The reason for this arises in the terms with $r_Z$ which reduce BR($ZZ$) and are not present for BR($\gamma \gamma$)  (see Eq.~(\ref{zz}) and (\ref{aa})). Finally, we observe that in the region of small $s_A$ and $C_H$ the $t\bar t$ decay channel starts to compete and emerges as the most sensitive one for both graviton masses. In fact, $t\bar t$ spreads across larger regions as $m_{X^*}$ increases since it is favored by phase space and a larger reconstruction efficiency for highly boosted top quarks but still it is far from reach at 8 TeV. On the other hand, the $b\bar b$ decay channel does not appear as the most sensitive one in any region of the parameter space. An explanation for this lies in two facts: the analysis includes $b\bar b$ annihilation as the only graviton production mechanism for this channel and, for increasing $m_{X^*}$, $b$-tagging is less efficient as the bottoms are more boosted. 

Up to now the analysis has made focus on the most sensitive (MS) channels. We  study next which channels present the more relevant subleading sensitivities in the allowed parameter space of the model. With this in mind, we have divided the region of the parameter space shown in the upper panels of Fig.~\ref{8tev-universal-leading} into a grid where we present the next to most sensitive (nMS) channels in each different section of that grid. This is displayed in the lower panels of Fig.~\ref{8tev-universal-leading} for the same two graviton masses, $m_{X^*}=0.5$ TeV and $m_{X^*}=1$ TeV. The numbers that we present correspond to the point in the center of each rectangle. For $t\bar t$ the values can fluctuate with the mixing angle of $t_L$ and $t_R$. 

Some observations are in order. We have shown only those decay channels with a ratio ${\cal S}_{\mbox{\scriptsize nMS}}/{\cal S}_{\mbox{\scriptsize MS}} > 0.1$, lower values are phenomenologically irrelevant. For both $m_{X^*}=0.5$ TeV and $m_{X^*}=1$ TeV, $WW$ is the next to the most sensitive channel in the region dominated by $ZZ$. We also see that $WW$ becomes more sensitive as $m_{X^*}$ increases because of a relative improvement in the $WW$ sensitivity with respect to the one of $ZZ$. Moreover, for $m_{X^*}=1$ TeV, $hh$ has a better sensitivity in the region with a relative large $C_H$ (the region where $ZZ$ is the most sensitive channel) compared to the case of $m_{X^*}=0.5$ TeV where it is negligible because of phase space suppression. In the region where the $\gamma \gamma$ channel is the most sensitive, the next one is $ZZ$ apart from a small region with low values of $C_H$ where $t\bar t$ takes its place, this occurs for both graviton masses. Finally, in the region defined by $s_A \lsim 0.1$ and $C_H \lsim 0.2$, the $ZZ$ channel is the next to the most sensitive one ($t\bar t$) for both graviton masses. Interestingly, for $m_{X^*}=1$ TeV and $C_H\gtrsim 0.4$, hh is not far from the most sensitive channel.

\subsubsection{Non-Universal couplings}

\begin{figure}
\begin{center}
\includegraphics[width=0.47\textwidth]{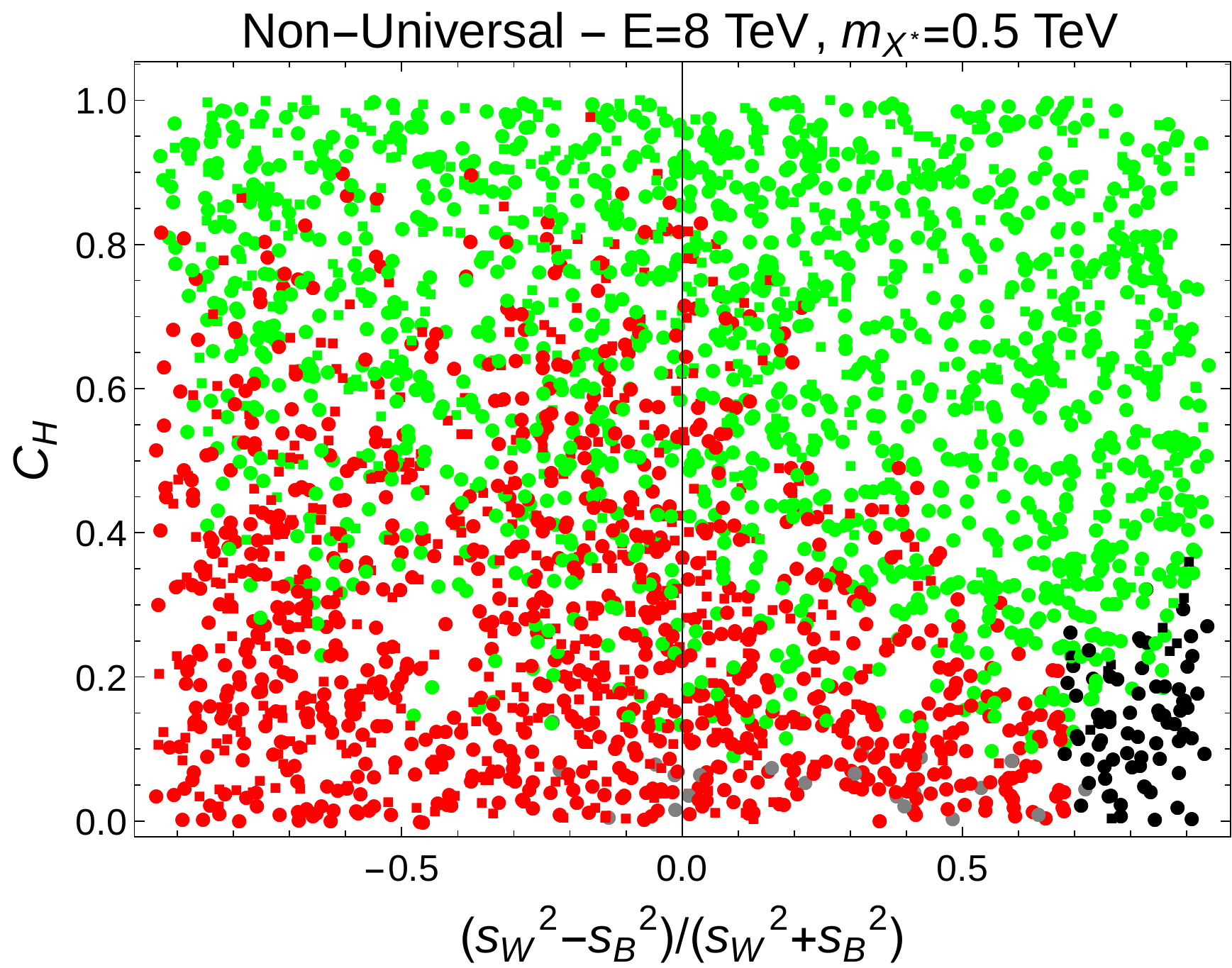}~
\includegraphics[width=0.47\textwidth]{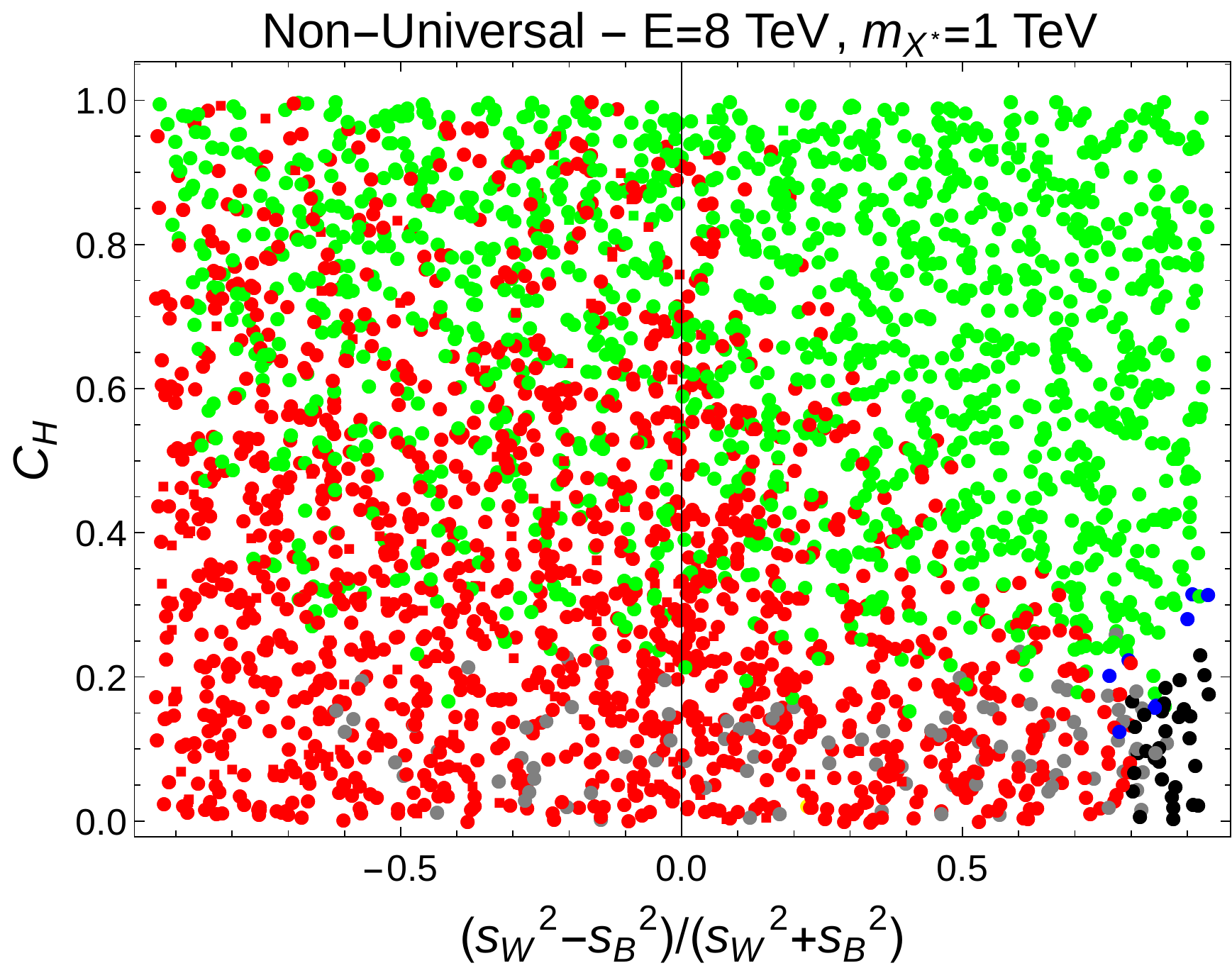}
\caption{
\small
Dominant channels in the plane $(s_W^2-s_B^2)^2/(s_W^2+s_B^2)^2$ vs. $C_H$ for LHC at 8 TeV. On the left we show the points that pass all the bounds for $m_{X^*}=0.5$ TeV and on the right for $m_{X^*}=1$ TeV. The colors indicate which channel has the largest ${\cal S}$: red for $\gamma\gamma$, green for $ZZ$, black for $Z\gamma$, gray for $t\bar t$ and blue for $WW$.
}
\label{8tev-nonuniversal}
\end{center}
\end{figure}

As discussed in sec.~\ref{sec:model}, for non-universal couplings $X^*$ can decay to $Z\gamma$. The strength of this channel is proportional to $(s_W^2-s_B^2)^2$. We have performed a random scan of the parameters, allowing different $s_B$ and $s_W$ in the range $(0.1,0.4)$. In Fig.~\ref{8tev-nonuniversal} we show, for the points that are not excluded by the bounds, the most sensitive channel (maximum ${\cal S}$) as function of  the non-universality and $C_H$ for LHC at 8 TeV. The left panel is for $m_{X^*}=0.5$ GeV, and the right one for $m_{X^*}=1$ TeV (see figure caption for the color encoding).

Let us describe first some common features of both masses. For large $C_H$ the $ZZ$ channel dominates, since in this regime the longitudinal $Z$ polarization has large couplings. For $s_W>s_B$, due to the smallness of the Weinberg angle, the $ZZ$ channel is favored over the $\gamma\gamma$ one. By similar reasons, for $s_B>s_W$, near the left border of the plots, the $\gamma\gamma$ channel dominates over $ZZ$. Near the right border and for small $C_H$, such that the longitudinal $ZZ$ polarization is suppressed, $Z\gamma$ dominates, since in that region the non-universal coupling is maximized. Also for small $C_H$, with only a mild dependence on the violation of universality, $t\bar t$ can sometimes dominate over the bosonic channels since the top reconstruction in Ref.~\cite{CMS:2016zte} is optimized for these graviton masses.

The dependence with $m_{X^*}$ can be studied by analyzing the differences between both figures. Roughly speaking the plots are very similar. For larger $m_{X^*}$ there are more points with $t\bar t$, there are at least two reasons for this effect: first the available phase space is far from threshold for $m_{X^*}=1$ TeV, second the limits of top pairs become more stringent than the limits of other channels. The latter is immediately seen in Table~\ref{sensitivities}, where the 8 TeV row of $t\bar t$ shows a ratio of sensitivity improvement roughly 30 \% better than in the $ZZ$ and $\gamma\gamma$ rows. One can also see that, for large $s_W$ and $m_{X^*}=1$ TeV, the channel $WW$ can dominate, whereas for $m_{X^*}=500$ GeV it does not. The difference arises from the sensitivity, with a better improvement in $WW$ than in $\gamma\gamma$ and $ZZ$.  

Although there are regions where just one of the channels dominates, between those regions there is an overlap where several channels can dominate. This happens because, although we only show explicitly the dependence on $(s_W^2-s_B^2)^2/(s_W^2+s_B^2)^2$ and $C_H$, to generate the set of points we have scanned over all the mixing and couplings, as explained at the beginning of this section. For universal couplings, $s_A$ and $C_H$ are the relevant parameters, and the dependence on the other parameters is negligible. For non-universal couplings, $s_A$ splits in different mixing for the different gauge groups. In particular $s_B$ and $s_W$ control the main decay channels, $VV$, such that for non-universal couplings a multi-dimensional plot as function of these mixing would be needed for a cleaner separation of regions.

\subsection{LHC at 13 TeV}\label{sec:13}

In this section we continue the analysis of the sensitivity of the graviton decay channels. In this case we consider data from LHC at 13 TeV and still quantify their sensitivity through the strength ${\cal S}$ and point out the most sensitive channels within the allowed parameter space of the model. We take into consideration the same two scenarios of the previous subsection: a graviton with universal or non-universal couplings. There is a difference with respect to the analysis performed with LHC at 8 TeV though.  As already discussed, the experimental searches at 13 TeV are not performed at the same luminosity. Therefore, in the following analysis we take a luminosity reference value of 13.3 fb$^{-1}$ which corresponds to three of those searches and we scale the remaining four, all carried out at slightly different luminosities, according to Eq.~\ref{lum}.

\subsubsection{Universal couplings}

We show in this section that $s_A$ and $C_H$ still control the degree of sensitivity of the different graviton decay channels at 13 TeV. We set the gravity scale $M_1$ to the same fixed value $M_1=3$ TeV. For relatively low graviton masses the couplings of a graviton to fermions ($s_{q}$, $s_{t}$ and $s_{b}$) do not play a significant role anywhere except for a modest region of the parameter space characterized by small values of $s_A$ and $C_H$; however, their impact increases for larger graviton masses as it can be seen in a larger overlap of colors, making of this behavior a qualitative difference with respect to the analysis carried out with LHC at 8 TeV. The reason for this is the improvement of top-tagging techniques for high $p_T$ top quarks ~\cite{CMS:2016zte}.

\begin{figure}[h!]
\begin{minipage}[c]{8cm}
\includegraphics[width=8cm]{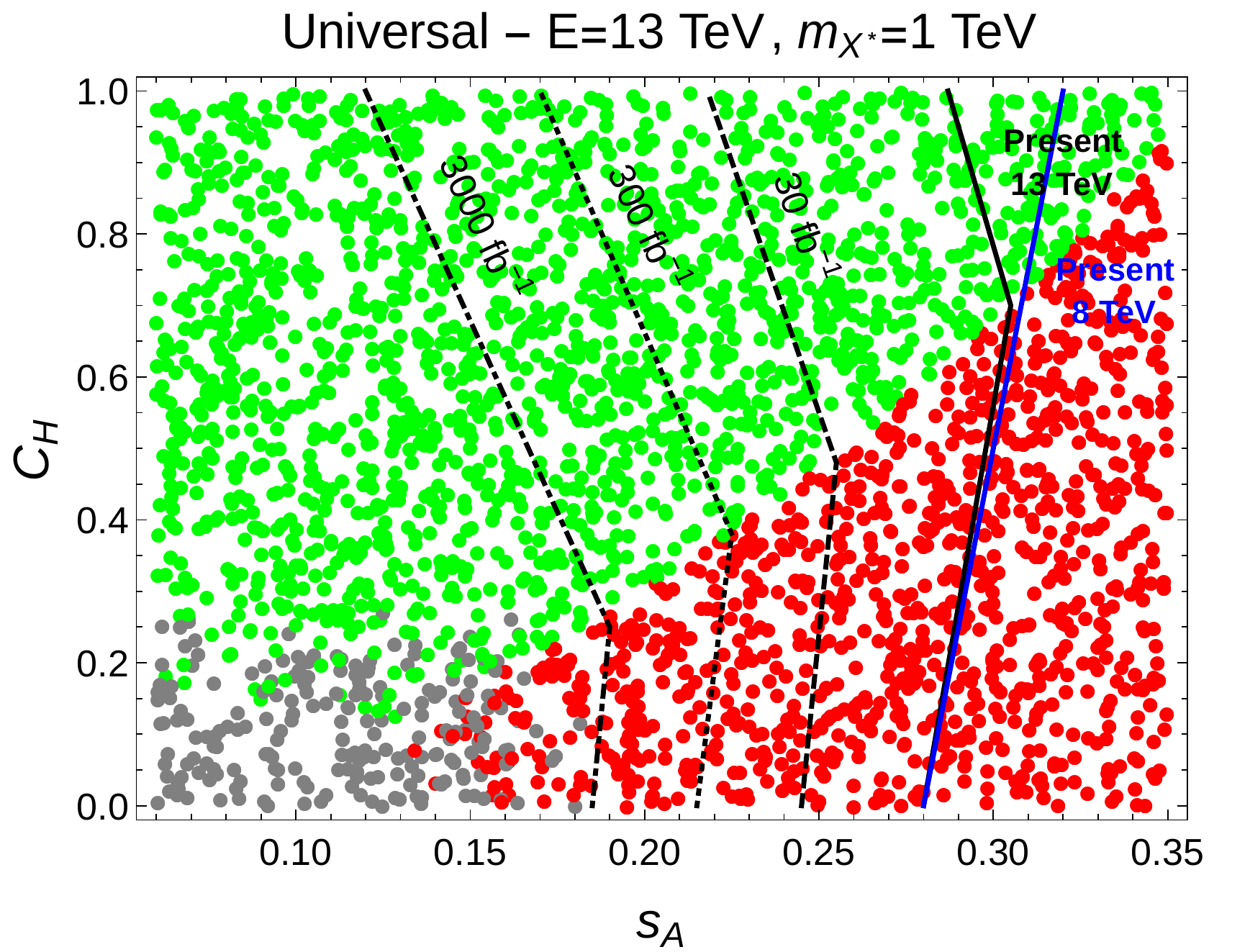}
\end{minipage}
\begin{minipage}[c]{8cm}
\includegraphics[width=8cm]{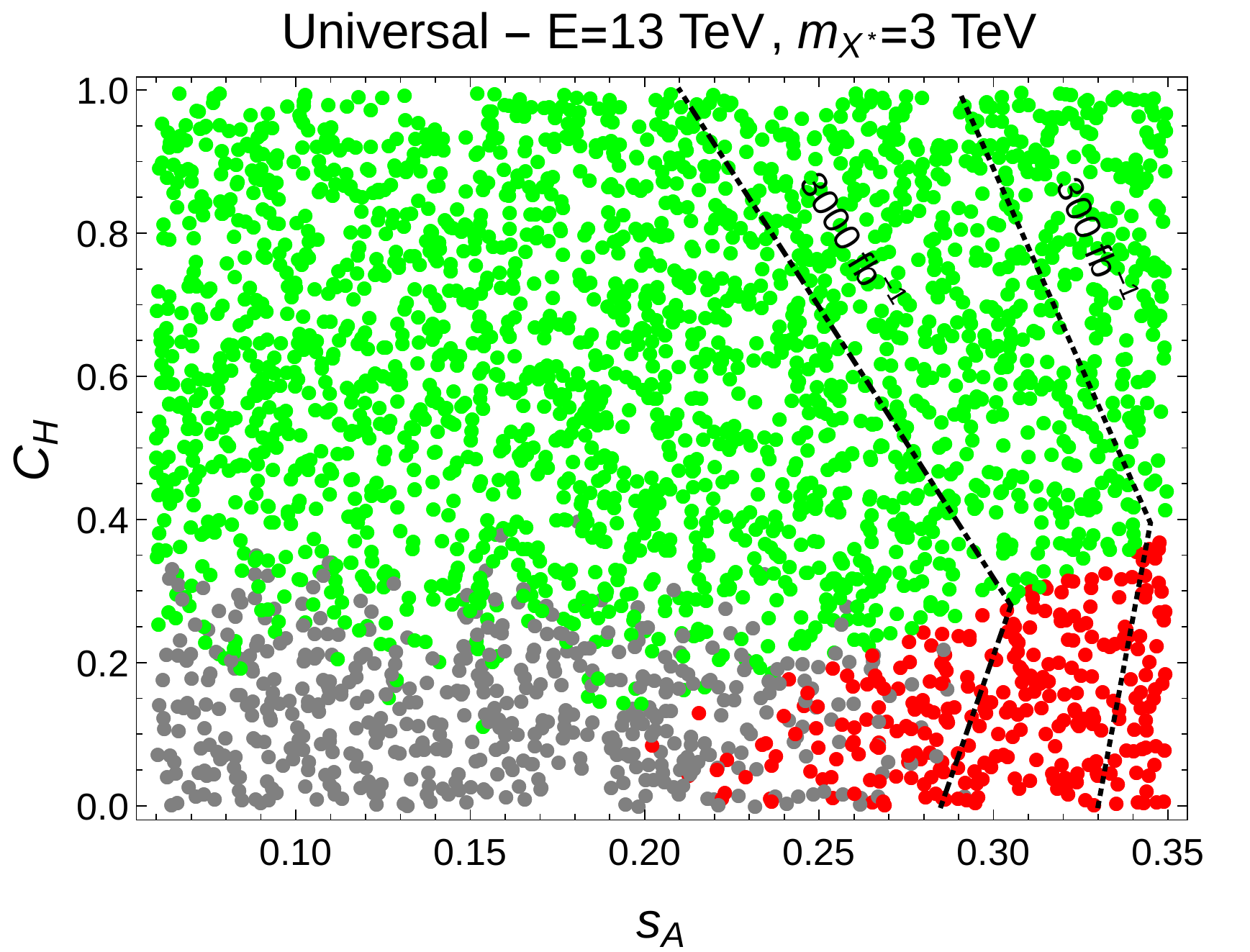}
\end{minipage}
\\
\begin{minipage}[c]{8cm}
\vspace*{0.5cm}
\includegraphics[width=8cm]{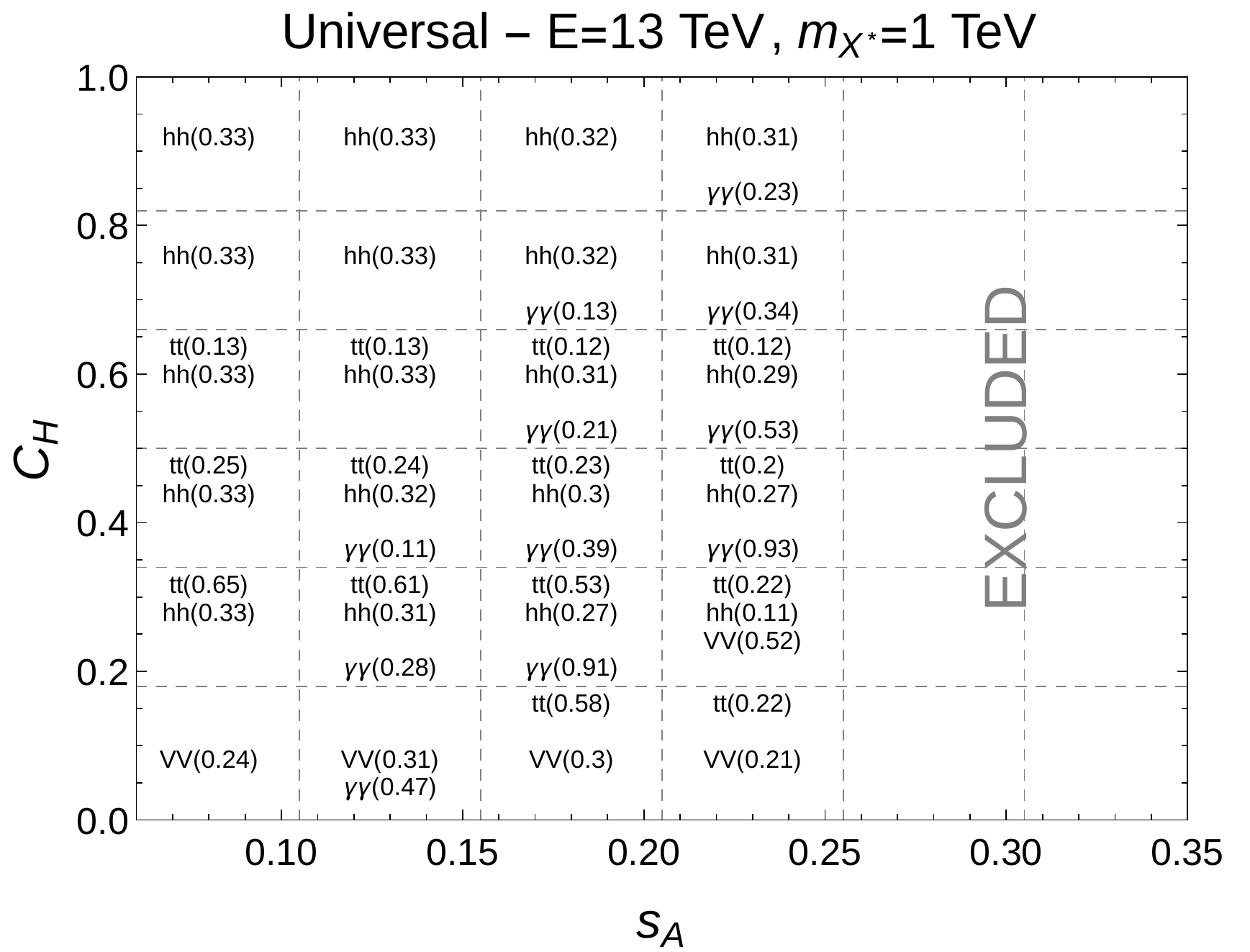}
\end{minipage}
\begin{minipage}[c]{8cm}
\vspace*{0.5cm}
\includegraphics[width=8cm]{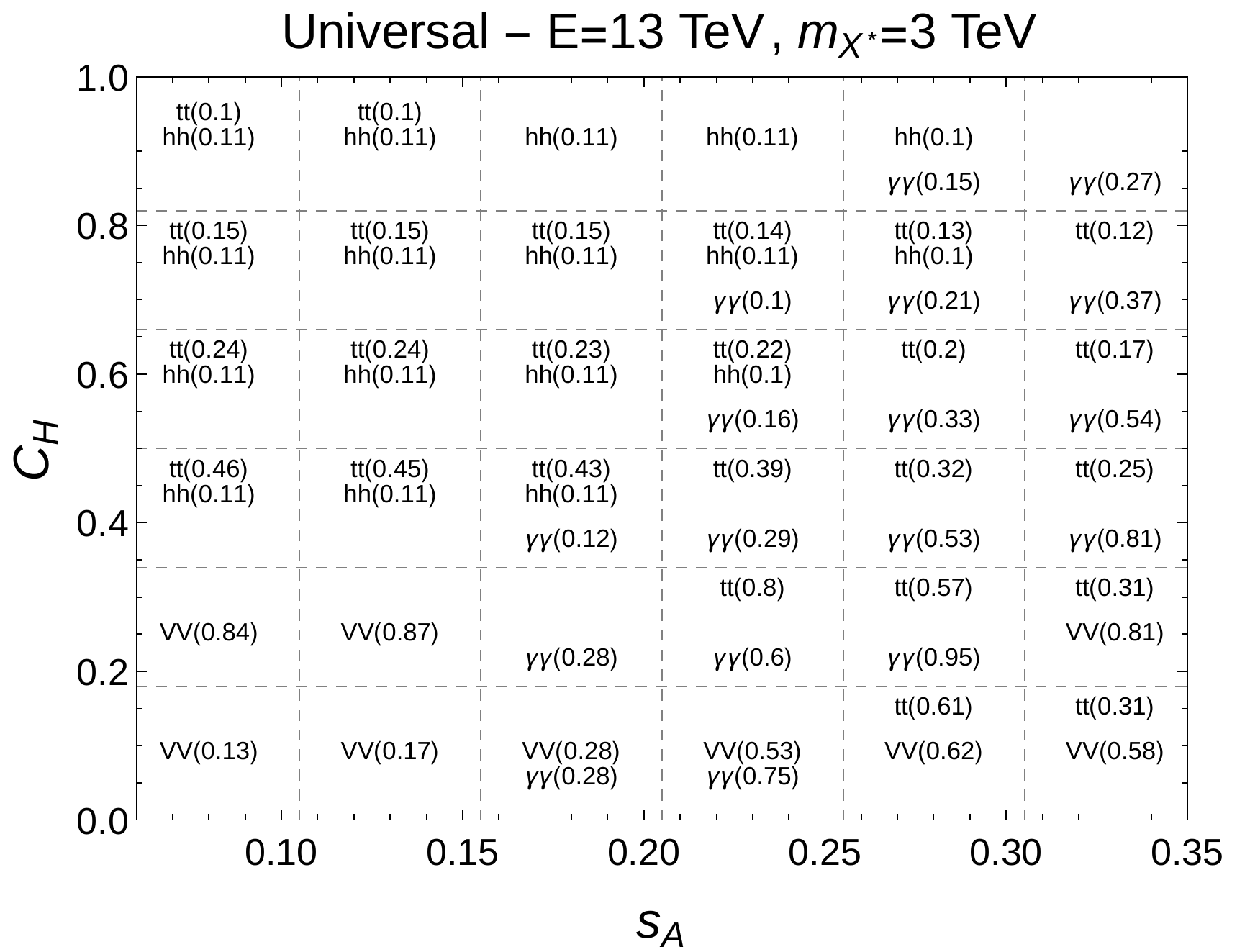}
\end{minipage}
\caption{
\small
Upper panels: The values of $C_H$ and $s_A$ for the points corresponding to the most sensitive graviton decays channels for LHC at 13 TeV within the scanned parameter space (defined in the text), for $m_{X^*}=1\text{ TeV}$ and $m_{X^*}=3\text{ TeV}$, and according to the following color coding: green, red and gray represent $ZZ$, $\gamma \gamma$ and $t\bar t$ channels, respectively. The black (blue) lines define exclusion regions for different luminosities: 30 fb$^{-1}$, 300 fb$^{-1}$, 3000 fb$^{-1}$ and present data at 13 TeV (8 TeV). Lower panels: Next to most sensitive graviton decay channels for LHC at 13 TeV, for $m_{X^*}=1\text{ TeV}$ and $m_{X^*}=3\text{ TeV}$. The numbers in parentheses next to a certain channel indicate its relative sensitivity (${\cal S}_{\mbox{\scriptsize nMS}}/{\cal S}_{\mbox{\scriptsize MS}}$) with respect to the most sensitive one for the central point of each rectangle of the grid.  
}
\label{13tev-universal-leading}
\end{figure}

Following the procedure presented in sec.~\ref{uni-8}, we generate once again scatter plots in the $s_A$-$C_H$ plane randomly scanning over all variable parameters within the same numerical ranges ($s_A \in (0.06,0.35),\ s_{q}\in (0.5,0.95),\ s_{t}\in (0.5,0.95),\ s_{b}\in (0.1,0.3)$ and $C_H\in (0,1)$) for two  graviton masses, $m_{X^*}=1$ TeV and $m_{X^*}=3$ TeV in this case. Likewise, each point in these plots indicates which graviton decay channel is the most sensitive one according to its value of ${\cal S}$. The results are shown in the upper panels of Fig.~\ref{13tev-universal-leading} and we now proceed to comment on several observations.

We maintain the color encoding used in the previous sections (see caption of Fig.~\ref{13tev-universal-leading}) where each color corresponds to a given decay channel. Note that green points stand now for the $VV$ channel, where $VV$ is $ZZ$ and $WW$ grouped together. The first observation concerns the plot for $m_{X^*}=1$ TeV which resembles the one for the same graviton mass at 8 TeV as it reveals a similar behavior regarding the strength ${\cal S}$, even when the plot at 13 TeV was obtained by means of scaling experimental limits at different luminosities. We observe then a comparable pattern in relation to the distribution of the most sensitive channels within the scanned parameter space. The most apparent difference in comparison with the plot for $m_{X^*}=1$ TeV at 8 TeV is related to the excluded region (${\cal S} > 1$) by present bounds~\footnote{The present bounds at 13 TeV have been obtained without making a rescaling of the luminosity but keeping the actual value used in the experimental analyses.}, now this region includes points corresponding to the $VV$ channel. Superimposing the present bounds at 8 TeV (blue line) on the left upper plot in Fig.~\ref{13tev-universal-leading},  we see that for small $C_H$ these are competitive to the ones at 13 TeV. However, as $C_H$ increases the present limits at 13 TeV exclude a larger region of the parameter space dominated by $VV$. This is a consequence of the behavior of the bounds for $m_{X^*}=1$ TeV at 13 TeV, which become more stringent than the 8 TeV ones because of the fact that both $WW$ and $ZZ$ in the final state can be reconstructed more efficiently as fat jets. These plots also show an estimation of  the projected limits at 300 fb$^{-1}$ and 3000 fb$^{-1}$. As expected, we first observe that for both luminosities the excluded parameter space region for $m_{X^*}=1$ TeV is larger than the corresponding one to $m_{X^*}=3$ TeV. Moreover, for $m_{X^*}=1$ TeV, 30 fb$^{-1}$ is already enough to exclude a considerable region but it is not even sufficient to start rejecting points for $m_{X^*}=3$ TeV. In particular, 300 fb$^{-1}$ would exclude the 50\% of the points in the near future and 3000 fb$^{-1}$ almost the 75\% in the long term. It is important to stress that these projections provide a general sense of how far is the value of ${\cal S}$ for a given point in the parameter space from ${\cal S}=1$. 

We also see that for each graviton mass the $VV$ channel becomes the most sensitive for relatively large values of $C_H$ as the longitudinal $W$ and $Z$ polarizations dominate whereas $\gamma \gamma$ has the best sensitivity in a region where $s_A$ is comparably large. This feature can be explained with the same arguments introduced in the sec.~\ref{uni-8}. We also recognize that $VV$ dominates over almost the whole parameter space for $C_H \gtrsim 0.2$ independently of the value of $s_A$ within the scanned range. This effect is a result of a significant relative enhancement in the exclusion power of $VV$ limits compared to $\gamma \gamma$ as $m_{X^*}$ increases, and it originates again in the fact that $WW$ and $ZZ$ each boson can emerge as one unique fat jet which is easier to identify for larger graviton masses. Within the region defined by low values of $s_A$ and $C_H$, where both $VV$ and $\gamma \gamma$ are relatively suppressed, the $t\bar t$ decay channel is the most sensitive. Its impact is larger as the graviton mass increases because of an increase of the available phase space and the larger reconstruction efficiency achieved for highly boosted top quarks, however, it is far from reach even for 3000 fb$^{-1}$ and for both graviton masses. This changes in next section where loop effects are taken into account. The $b\bar b$ channel is not competitive even in the left corner region since only $b\bar b$ annihilation produces this final state and the outgoing bottom quarks are too boosted to be tagged efficiently for the graviton masses we have explored at 13 TeV. Moreover, compared to $m_{X^*}=1$ TeV, we notice that there are not excluded points for $m_{X^*}=3$ TeV since the present bounds are still weak for this mass at 13 TeV.

We now analyze which channels are the next to the most sensitive ones throughout the allowed parameter space of the model. As in the case of LHC at 8 TeV, we divide  the scanned region of the parameter space shown in the upper panels of Fig.~\ref{13tev-universal-leading} into a grid. We then display in the lower panels of Fig.~\ref{13tev-universal-leading} the next to most sensitive  channels in each different section of that grid for $m_{X^*}=1$ TeV and $m_{X^*}=3$ TeV. The numbers in the grids correspond to the central point of each rectangle, and for the $t\bar t$ channel the values can fluctuate with the mixing angle of $t_L$ and $t_R$.

Regarding the lower panels of Fig.~\ref{13tev-universal-leading} we have the following observations. As in the case of LHC at 8 TeV, we have shown only decay channels with a ratio ${\cal S}_{\mbox{\scriptsize nMS}}/{\cal S}_{\mbox{\scriptsize MS}} > 0.1$. For $m_{X^*}=1$ TeV, in the region where $VV$ is the most sensitive channel, $hh$ is the next one for relatively large values of $C_H$ whereas $\gamma \gamma$ increases its impact as $C_H$ decreases and $s_A$ rises. For small $C_H$ and $s_A$ within the same $VV$ dominance region, $t\bar t$ becomes the next to the most sensitive decay channel. On the other hand, in the region dominated by $\gamma \gamma$, $VV$ is the next to the most sensitive channel for moderate values of $C_H$ and $s_A$, and $t\bar t$ for lower $s_A$. Regarding the region defined by $s_A \lsim 0.1$ and $C_H \lsim 0.15$ where $t\bar t$ is dominant, $VV$ is the next one and then $\gamma \gamma$ as $s_A$ grows. Therefore, we observe a similar pattern compared to the distribution of the next to most sensitive channels obtained for $m_{X^*}=1$ TeV at 8 TeV. In the case of $m_{X^*}=3$ TeV and in the region dominated by $VV$, $hh$ is still the next to the most sensitive channel for $s_A \lsim 0.25$ and $C_H \gtrsim 0.8$ but its effect is milder with respect to $m_{X^*}=1$ TeV. Yet in the region where $VV$ is the most sensitive channel and $C_H \lsim 0.8$, $\gamma \gamma$ ($t\bar t$) surpasses $hh$ for larger (lower) values of $s_A$. For $C_H \lsim 0.4$ and $s_A \gtrsim 0.25$, where $\gamma \gamma$ appears as the most sensitive, $VV$ and $t\bar t$ are competitive as next to most sensitive channels. Finally, for the region dominated by the $t\bar t$ channel corresponding to $C_H \lsim 0.4$ and $s_A \lsim 0.25$, $VV$ is the next to the most sensitive one and $\gamma \gamma$ starts to compete for larger $s_A$.

\subsubsection{Non-Universal couplings}

\begin{figure}[h!]
\begin{center}
\includegraphics[width=0.47\textwidth]{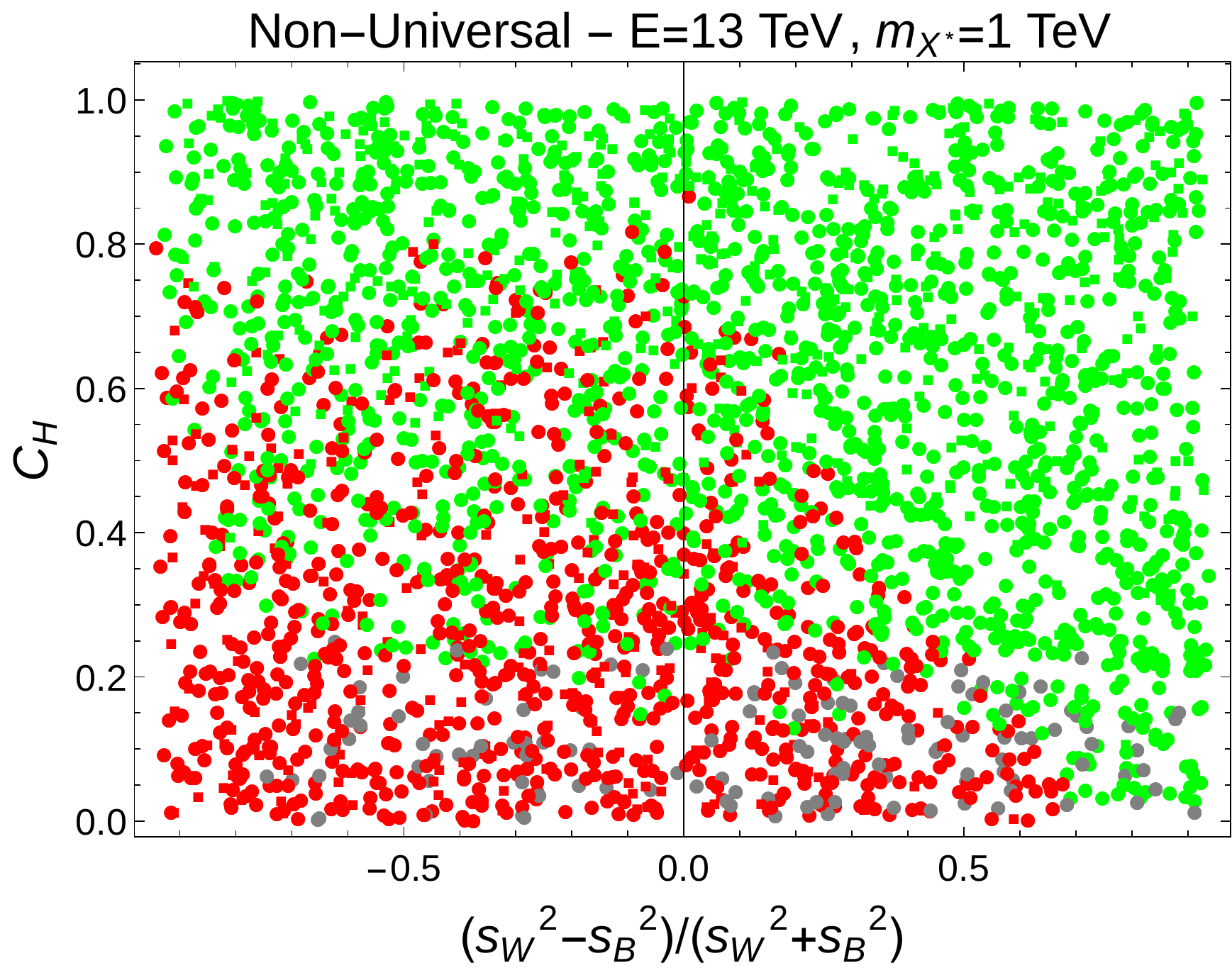}~
\includegraphics[width=0.47\textwidth]{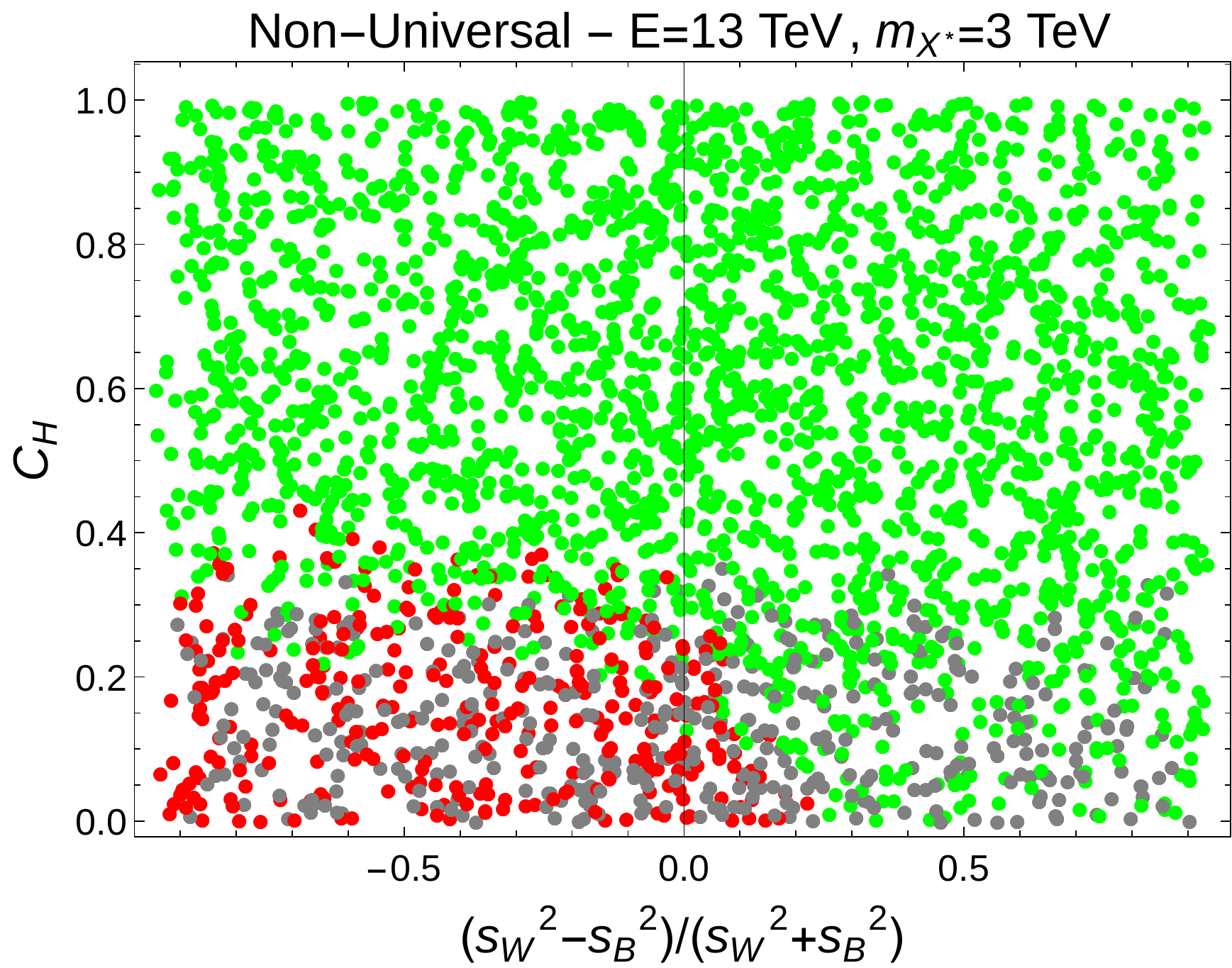}
\caption{
\small
Dominant channels in the plane $(s_W^2-s_B^2)^2/(s_W^2+s_B^2)^2$ vs. $C_H$ for LHC at 13 TeV. We only show the points that pass the bounds on ${\cal S}$ at this center of mass energy. On the left we show the results for $m_{X^*}=1$ TeV and on the right for $m_{X^*}=3$ TeV. The colors indicate which channel has the largest ${\cal S}$: red for $\gamma\gamma$, green for $ZZ$ and gray for $t\bar t$.
}
\label{13tev-nonuniversal}
\end{center}
\end{figure}

We have also analyzed ${\cal S}$ for LHC at 13 TeV with non-universal couplings. In Fig.~\ref{13tev-nonuniversal} we show our results for $m_{X^*}=1$ TeV and $m_{X^*}=3$ TeV, scanning randomly over the mixing and couplings, and for points not excluded by the bounds at 13 TeV only since we do not expect qualitative differences from those at 8 TeV.

We start analyzing $m_{X^*}=1$ TeV. Similar to 8 TeV, larger $C_H$ and $s_W$ favor the $VV$ channel (see figure caption for the color encoding). For small $C_H$, the coupling to longitudinal $V$ polarizations is suppressed, besides larger $s_B$ increases the coupling to $\gamma$, thus $\gamma\gamma$ dominates in the lower left region. Also for low $C_H$ the $t\bar t$ channel can sometimes become dominant. Although it is not shown in the figure, $Z\gamma$ is near maximal for the down-right corner, in fact if the limits of $Z\gamma$ are improved by a factor $\sim 1.7$ with the other limits remaining constant, black dots corresponding to this channel would appear for small $C_H$ and large positive $s_W^2-s_B^2$. The limits of the $VV$ channel are improved faster than the limits of $\gamma\gamma$ with increasing $m_{X^*}$, for that reason in the case of $m_{X^*}=3$ TeV $VV$ gains over $\gamma\gamma$ except for the lower left corner, where the $V$-coupling is suppressed. In this case $Z\gamma$ is far from maximal strength, an improvement of one order of magnitude of ${\cal S}$ in this channel without an improvement of the other channels is required to obtain $Z\gamma$ as the dominant one.

\subsection{Discussion}
\label{sec:predictions}

We now compare the results obtained in the previous subsections for LHC at 8 and 13 TeV, discuss some general phenomenological features of a massive graviton, and comment further on the relevance of the strength ${\cal S}$.

Beginning with the case of universal couplings, we first mention that the electroweak bosonic graviton decay channels are the most sensitive (in the sense of reaching a maximum of ${\cal S}$) over almost the whole parameter space for both center of mass energies. The degree of sensitivity is mainly controlled by two parameters, $C_H$ and $s_A$, and the weak channels, $ZZ$ and $VV$ at 8 and 13 TeV respectively, tend to dominate for large values $C_H$, whereas $\gamma \gamma$ is dominant for large $s_A$ provided that $C_H$ is small enough. This pattern is observed for the three graviton masses we have analyzed: 0.5, 1 and 3 TeV. The impact of the weak channels becomes larger with rising energy covering a very large proportion of the parameter space ($C_H \gtrsim 0.2$) for $m_{X^*}=3$ TeV at 13 TeV. In the region where $C_H$ and $s_A$ are small, the $t\bar t$ channel is the most sensitive one and its sensitivity improves with an increase of both graviton mass and energy. This tendency is fulfilled for $m_{X^*}=3$ TeV where $t\bar t$ is the most sensitive channel for $s_A \lsim 0.2$ (and $C_H \lsim 0.2$) exceeding $\gamma \gamma$. Concerning the present bounds, the strongest one is given for $m_{X^*}=0.5$ TeV at 8 TeV excluding $s_A \gtrsim (0.23-0.26)$ depending on the values of $C_H$. For larger graviton masses, the bounds are weaker, as expected, and in particular for $m_{X^*}=1$ TeV the present bounds at 8 TeV and 13 TeV are competitive at low values of $C_H$ and the latter prevails at large values of this parameter. Moreover, combined together they exclude only a slightly smaller region than the present bound for $m_{X^*}=0.5$ TeV at 8 TeV. Regarding the next to most sensitive channels, we find a similar pattern at 8 and 13 TeV regarding the dependence on couplings and graviton masses, and their main characteristics have been already discussed previously. Finally, in the case of non-universal couplings, we see that the main difference between 8 and 13 TeV is a relative worsening in the sensitivity of the $Z \gamma$ channel at 13 TeV. Finally, an observation regarding the gravity scale $M_1$. All along this work we have fixed it to a conservative value of 3 TeV. Since it plays the role of a global normalization factor just affecting the graviton production cross section, the only effect of changing its value is a rearrangement of reaches at different luminosities. In particular, decreasing $M_1$ leads to a displacement of the curves towards smaller $s_A$ values giving place to a larger constraint of the parameter space.

Besides the discussion on the most sensitive channels presented in the previous paragraphs, we should also address some points which should help to distinguish the theoretical framework presented in this work from others in case of an eventual excess. Some distinctive features which should be noticed in case of a massive graviton is discovered.  

Since the graviton is mainly created by gluon fusion, a generic characteristic of the spin 2 nature of the graviton is the presence of forward photons in the final state. Another interesting feature is found from Eqs.~(\ref{ff}) and (\ref{za2}). It is possible to solve for one partial width as a function of the others. The exact solutions are too long, but in the limit with $r_{Z,h}\to 0$ one obtains
\begin{equation}
 \frac{ 8 \sigma_{Z\gamma} \, \sigma_{\gamma\gamma}}{\tan^2(2\theta_w) } = \left( \sigma_{ZZ} - \sigma_{\gamma\gamma} - \sigma_{hh} - \frac{2 \sigma_{Z\gamma}}{\tan^2(2\theta_w)} \right)^2 ,
\label{prediction}
\end{equation}
where $\sigma_{ij} = \sigma(pp\to X^*)\, \times \, \mbox{BR}(X^* \to ij)$.  By replacing $\sigma_{Z\gamma} \to 0$ one obtains the universal-couplings case, which is rather simple and, as expected, corresponds to the sum of the transversal and longitudinal polarizations of the $Z$. This result is a consequence of the graviton being neutral under the SM gauge group and of the limit that we have considered: $X^*$ much heavier than the SM particles.

Regarding the strength ${\cal S}$ it is important to emphasize the following observation. In re-analyzing Fig.~\ref{brs} we can see that there are channels with large branching ratios for a massive graviton, but the experimental sensitivity fails to take full profit of them. These are mainly $t\bar t$ and $b \bar b$. Boosted-top reconstruction has received very important advancements in the last years, however, this is still not true for $b$-tagging.  Although it seems a tough job, plots like the ones in Fig.~\ref{brs} would indicate the effort in this direction could be important to increase sensitivity to a massive graviton.

A final remark in order to close the general phenomenological picture of the theoretical framework proposed here. It may be timely to remind the reader that the phenomenology of a pNBG Higgs has been largely studied through the analysis of new light fermions states with masses of order TeV (custodians) which may be produced and detected at LHC~\cite{Contino:2008hi,DeSimone:2012fs}, and also through double Higgs production~\cite{Contino:2010mh,Contino:2012xk}.

\section{Stability upon loop corrections}\label{sec:loop}
The predictions presented in the previous sections were obtained via tree-level calculations.
In order to test the stability of our results, we evaluated the impact of the fermion loop contributions to the production of the massive graviton and its decay into gluons and photons, closely following the discussion given by us in Ref.~\cite{Alvarez:2016uzm}.
To this end, we included contributions from bottom and top-quark loops, and also from the heavy partners of all the fermions of the SM.
Other decay channels, as well as the small contribution to the production cross section coming from the $b\bar{b}$ initial state, are still evaluated at tree-level.

We start by analyzing the loop corrections to the $\gamma\gamma$ decay.
Working in dimensional regularization with $D=4-\epsilon$, and following the calculations performed in Ref.~\cite{Geng:2016xin}, we can account for the sum of the tree-level and loop-induced contributions by modifying the bare coupling in the following way,
\begin{equation}
C_\gamma^0 \to C_\gamma^0 + \f{\alpha}{2\pi}\sum_i Q_i^2 C_i \f{N_c}{3} \left(
\f{2}{\epsilon}-\gamma_E+\ln(4\pi)+
A_G(\tau_i,\mu) 
\right)\,,
\end{equation}
where the sum includes all the fermions that couple to the massive graviton, and has two different contributions for the right and left-handed chiralities.
Here $Q_i$ represents the charge of the fermion (in units of the positron charge), $C_i$ stands for its coupling to the graviton and $N_c$ for the number of colors.
The Wilson coefficient can be renormalized as
\begin{equation}
C_\gamma = C_\gamma^0
+ \f{\alpha}{2\pi}\sum_i Q_i^2 C_i \f{N_c}{3} \left(
\f{2}{\epsilon}-\gamma_E+\ln (4\pi) 
\right)\,,
\end{equation}
which leads to the following effective coupling
\begin{equation}
C_\gamma^{\text{eff}}(\mu) =
C_\gamma + 
\f{\alpha}{2\pi}\sum_i Q_i^2 C_i \f{N_c}{3}
A_G(\tau_i,\mu)\,.
\end{equation}

In the case of SM fermions, for which the variable $\tau_i = 4 m_i^2/m_{X^*}^2$ is always lower than 1, the function $A_G$ can be written in the following way  \cite{Geng:2016xin}
\begin{eqnarray}
A_G(\tau,\mu) &=&  -\f{1}{12}\bigg[
-\f{9}{4}\tau(\tau+2) [2 \tanh^{-1}(\sqrt{1-\tau}) - i\pi]^2
\\
&+& 3(5\tau+4)\sqrt{1-\tau}[2 \tanh^{-1}(\sqrt{1-\tau}) - i\pi]
- 39\tau - 35 -12 \ln\f{\mu^2}{m_i^2} \bigg]\,.\nonumber
\end{eqnarray}
In order to obtain the corresponding phenomenological results, we set the renormalization scale to the value $\mu = m_{X^*}$, which represents the energy scale of the process.
It is worth mentioning that this choice warranties a finite result for the loop function $A_G$ in the limit $\tau\to 0$.

On the other hand, for the heavy partners of the SM fermions we always have $\tau > 1$.
The function $A_G$ in this case simply reads
\begin{eqnarray}
A_G(\tau,\mu) &=& -\f{1}{12}\bigg[
\f{9}{4}\tau(\tau+2) [2 \tan^{-1}(\sqrt{\tau-1}) - \pi]^2
\\
&-& 3(5\tau+4)\sqrt{\tau-1}[2 \tan^{-1}(\sqrt{\tau-1}) - \pi]
- 39\tau  - 35 -12 \ln\f{\mu^2}{m_i^2} \bigg]\,.\nonumber
\end{eqnarray}
However, the appropriate scale choice is different from the previous one. 
The value of the scale $\mu$ is imposed in this case by a matching condition between the effective theory under consideration, valid up to a certain cut-off $\Lambda \sim M_1$, and the full theory, which is of course also valid above this scale.
Within this framework, in order to avoid large logarithms in the Wilson coefficients, as it is the usual procedure, the renormalization scale is set at the mass of the heavy fermions.
We do not include the effects coming from the running
from $\mu = m_{\psi_1} \sim {\cal O}(\text{TeV})$ to $\mu = m_{X^*}$ since they are expected to be small.

The features described above for the photon case also apply for the loop-induced contributions to the coupling between the massive graviton and a pair of gluons, and the resulting effective coupling therefore can be written as
\begin{equation}
C_g^{\text{eff}}(\mu) = C_g 
+ 
\f{\alpha_S}{2\pi}\f{1}{6}\sum_i C_i 
A_G(\tau_i,\mu)\,,
\end{equation}
where the sum runs over each chirality of the fermions that carry color.

Regarding the contributions from the heavy partners of the SM fermions, we have studied different embeddings.
We have considered MCHM$_5$, in which four composite quarks are introduced for each generation, in the following representations of SO(5): $q^u_1,u_1\sim{\bf5}_{2/3}$ and $q^d_1,d_1\sim{\bf5}_{-1/3}$, while for the leptons we used $L_1,e_1\sim{\bf5}_{-1}$.
On the other hand, we also studied MCHM$_{10}$, with $q_1,u_1,d_1\sim{\bf10}_{2/3}$ and $L_1,e_1\sim{\bf10}_{-1}$~\cite{Contino:2006qr}. We have also considered a set of representations which allows to solve the deviation in $A_{FB}^b$~\cite{Ciuchini:2014dea}, by embedding $q^d_1\sim{\bf16}_{-5/6}$ and $d_1\sim{\bf4}_{-5/6}$~\cite{Andres:2015oqa}.

We have evaluated separately the effect arising from the SM fermions and their heavy partners, for different graviton masses and always for the case of universal couplings.
In all cases, we found that the largest correction comes from the bottom and top-quark contributions, while the effect due to the inclusion of heavy partners in the loops is clearly subleading, due to the suppression induced by the loop function $A_G$ for large masses.
More precisely, the most important deviation from the tree-level results occurs in the effective coupling between the massive graviton and a pair of gluons; the reason for that is simply the larger value of the strong coupling constant $\alpha_S$ compared with the QED coupling $\alpha$ entering in the corrections to the diphoton decay.

Given that the decay into two gluons is phenomenologically not relevant, our tree-level results for the most sensitive channel and subleading channels remain largely unchanged when the loop-induced contributions are included.
However, the bottom and top-quark loop contributions became relevant for the production of the massive graviton, specially for relatively small values of the tree-level coupling $s_A$, generating an increase in the total cross section.
This, in turn, produces a small modification in the currently allowed region of the parameter space, while larger corrections are found for the reach estimate, depending on the values of $s_A$ under consideration.

\begin{figure}
\begin{minipage}[c]{8cm}
\includegraphics[width=8cm]{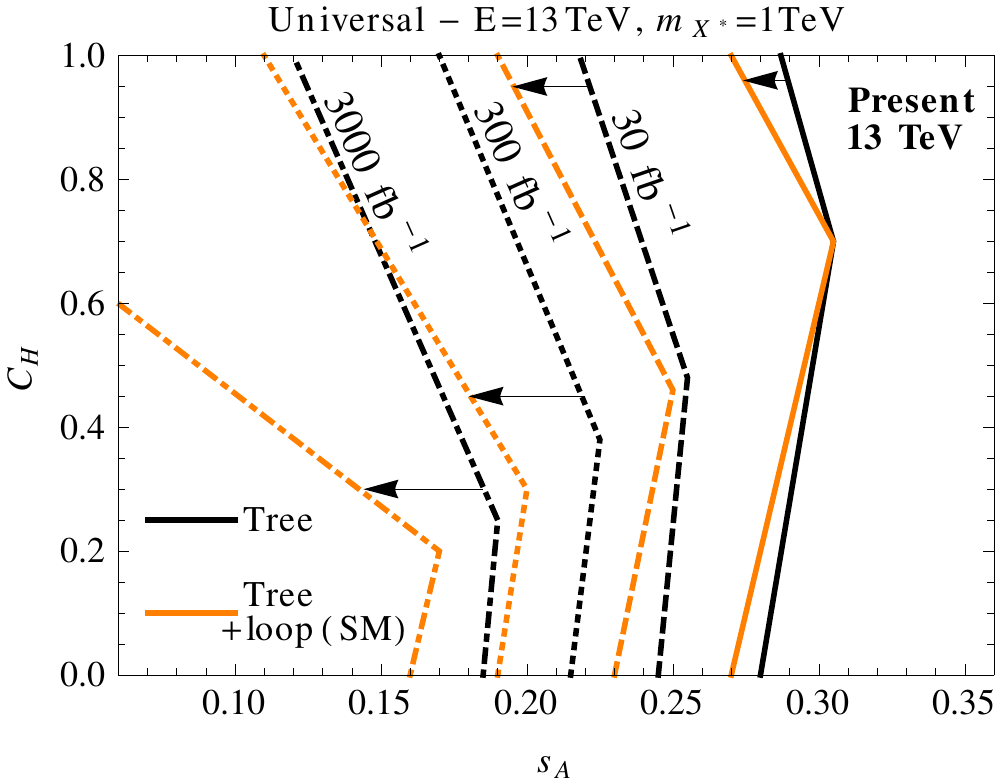}
\end{minipage}
\begin{minipage}[c]{8cm}
\includegraphics[width=8cm]{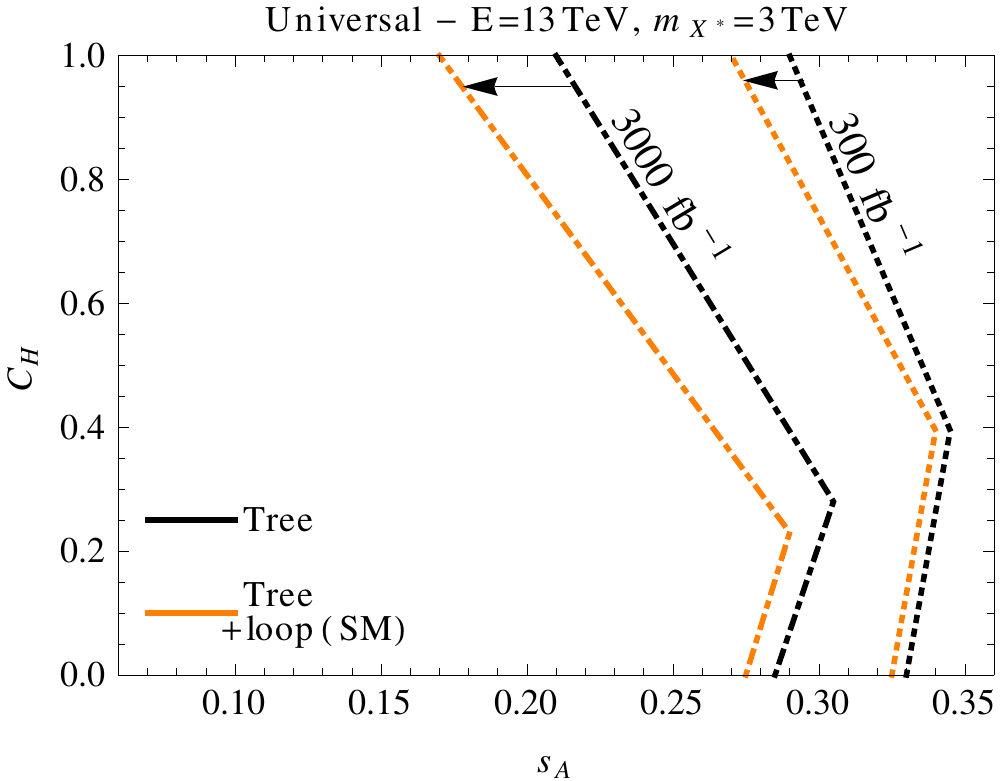}
\end{minipage}
\caption{
\small
Exclusion regions for different luminosities, obtained at tree-level (black) and including loop effects (orange), for $m_{X^*}=1\text{ TeV}$ (left) and $3\text{ TeV}$ (right).
}
\label{fig:loop}
\end{figure}

In Fig. \ref{fig:loop} we show the exclusion regions for different luminosities at $13\text{ TeV}$ for $m_{X^*}=1$ and $3\text{ TeV}$.
We present the tree-level results, which were already included in Fig. \ref{13tev-universal-leading}, and the predictions including the loop contributions from SM fermions.
We can observe clear modifications in the curves, always finding more stringent bounds in the parameter space once the loop effects are included, given that, as was stated before, the SM loop-induced contributions generate an increase in the production cross section.
As expected, the difference becomes larger as the value of $s_A$ decreases, where the relative size of the loop contribution is larger.
For the $m_{X^*}=1\text{ TeV}$ case, the most relevant modifications occur for the 300 fb$^{-1}$ and 3000 fb$^{-1}$ curves, which after the inclusion of the loop effects leave a substantially smaller part of the parameter space unexplored.
On the other hand, for $m_{X^*}=3\text{ TeV}$ the corrections are more moderate, essentially because the regions which can be explored with luminosities up to 3000 fb$^{-1}$ are still dominated by the tree-level production mechanism.

Of course, a consistent approach would also include the  corrections coming from fermion loops for the rest of the decay modes, in particular to $Z$ and $W$ bosons.
This is beyond the scope of this work.
However, from the small size of the contributions that we found for the diphoton decay channel, it is reasonable to expect that these loop-induced corrections, also of electroweak nature, will also be subleading, suggesting the stability of the results presented in this work.

\section{Conclusions}\label{sec:conclusions}

Along this work we have presented a consistent picture for a massive spin two state interacting with the SM fermions and gauge fields, as well as a pseudo Nambu-Goldstone boson Higgs, and we have studied its phenomenology at the LHC. The picture we presented can be extended to warped/composite scenarios, where the massive graviton corresponds to the first KK mode, and the Higgs to the fifth component of a five dimensional gauge field. 

We have described the massive graviton in the framework of a 2-site model including partners for fermions and gauge bosons. The resonances mix with the elementary states through mixing angles that measure the degree of compositeness of the mass eigenstates. One of the key features is that the graviton-Higgs interaction is modulated by a parameter $0\leq C_H\leq 1$ whose natural value can be estimated around $0.2 \lesssim C_H \lesssim 0.8$, in contrast to $C_H=1$ for a composite Higgs.  The main consequences of this variable in the model phenomenology is that it could suppress the graviton decay to $hh$ and to weak gauge bosons through its longitudinal polarizations.

To study the phenomenology of the model we have pursued the question of which channel is the most sensitive one to find a signal as a function of the parameters of the theory.  We have varied the graviton mass from $0.5$ to $3$ TeV and the mixing angles for the gauge bosons and third generation fermions, as well as $C_H$. We have considered separately the cases of universal and non-universal gauge bosons couplings. 

Instead of designing a search strategy and computing a reach analysis for each channel, we have defined the strength (${\cal S}$) of each channel as the ratio of the predicted signal to the experimental limits at the 95\% CL in available ATLAS and CMS searches for the corresponding channel.  We have centered the discussion on which channels would be the most sensitive ones by comparing ${\cal S}$ for all channels in different points of the parameter space and different energy and luminosity scenarios.  We have shown that addressing the question about which channel is the most sensitive within this framework is simple and provides an important insight into the model phenomenology.  Moreover, we have discussed how this way of addressing the question includes in a simple manner many important experimental aspects, as for instance how a given channel may undergo drastic changes in its sensitivity due to a modification in the way of reconstructing or detecting particles as a function of their energy.  An example of this feature is how weak gauge bosons are reconstructed or detected in leptonic channels at low energies and in hadronic channels as fat jets at higher energies.

For low and medium mass graviton ($m_{X^*} \approx 0.5$--$1$ TeV) we find that $ZZ$ and $\gamma\gamma$ channel dominate the most sensitive channel for most of the parameter space.  In the universal couplings case $ZZ$ is preferred for large $C_H$ and $\gamma\gamma$ for large gauge boson mixing angle $s_A$.  When both $C_h$ and $s_A$ are small, $t \bar t$ is preferred, however far from reach.  A similar situation holds for non-universal couplings with the addition that $Z\gamma$ could be the most sensitive channel for large departures of $s_W$ from $s_B$.  See Figs.~\ref{8tev-universal-leading} and \ref{8tev-nonuniversal} for details.

For large mass graviton ($m_{X^*} \approx 3$ TeV) we find that $ZZ/WW$  increase its domination as the most sensitive channel in a larger region in parameter space, except in the case where $C_H$ becomes quite small, where $\gamma\gamma$ and $t\bar t$ dominate for large and small $s_A$, respectively.  Again, where $t\bar t$ dominates, the signal is so weak that is far from reach.  These results are depicted in Figs.~\ref{13tev-universal-leading} and \ref{13tev-nonuniversal}.

We have also studied the subdominant channels in all cases.  We found that in general $ZZ$ and $\gamma\gamma$ are also mutually subdominant in most of the parameter space. In addition, $hh$ may be also found for large $C_H$, and $WW$ is always behind $ZZ$. The $t\bar t$ channel shows up as $C_H$ and $s_A$ do not take large values, whereas $b\bar b$ and $gg$ are never found within the studied leading subdominant channels since they are always below the $10\%$ of the dominant channel.

Summarizing the model phenomenology, we find that $ZZ$, $WW$, and $\gamma\gamma$ are the best channels to look for a signal. The former becoming more important as graviton mass increases.

We have also studied expected features of a massive graviton which could help to distinguish it from other particles in case an excess is found in the future.  These include, for instance, a relationship between the branching ratios in different channels.

As a complement of the previous analysis, and since 2-site models --and warped/composite scenarios-- predict numerous Kaluza-Klein fermion partners depending on the embedding, we have estimated the size of the 1-loop contributions arising from these new particles.  We have found that even for large embeddings the loop suppression makes these corrections not important in the currently allowed region in parameter space nor in the graviton branching ratios.  However, since modifications in the graviton to gluon coupling due to loop effects can be relatively important when the tree level coupling is small, we find moderate corrections in the reach estimate in this case. 

Finally, we have assumed a simple statistic regime for the scaling of the strength ${\cal S}$ and we have estimated the discovery reach for LHC at 30 fb$^{-1}$, 300 fb$^{-1}$ and 3000 fb$^{-1}$. We found that for graviton masses of up to $\sim 1$ TeV a rather large fraction of the parameter space should be explored with 300 fb$^{-1}$ and a major area with 3000 fb$^{-1}$. On the other hand, a large region would remain unexplored for graviton masses of $\sim 3$ TeV and beyond, where improved search strategies could be valuable.


\section*{Acknowledgments}

We thank Da Huang and Gonzalo Torroba for discussions on the radiative corrections, as well as An\'ibal Medina for useful discussions. E.A. thanks participants in ``Voyages Beyond the SM'' workshop for motivating environment and discussions in the course of this work. L.D. thanks ICAS for hospitality during the completion of this work. This work was partially supported by ANPCyT PICT 2013-2266.

\bibliography{biblio}

\end{document}